# AN APPROACH TO REMOVE KEY ESCROW PROBLEM IN ID-BASED ENCRYPTION FROM PAIRING

*Dissertation submitted to Jawaharlal Nehru University*
*in partial fulfilment of the requirements*
*for the award of the degree of*

**MASTER OF TECHNOLOGY**

**IN**

**COMPUTER SCIENCE AND TECHNOLOGY**

**BY**

**MAHENDER KUMAR**

**13/10/MT/016**

Under

the Supervision of

**Prof. C. P. KATTI**

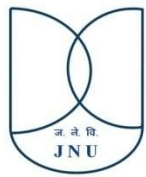

**SCHOOL OF COMPUTER & SYSTEMS SCIENCES**
**JAWAHARLAL NEHRU UNIVERSITY**
**NEW DELHI-110067**
**INDIA**
**2015**

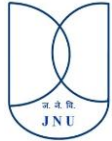

# CERTIFICATE

This is to certify that the dissertation entitled **"An Approach to Remove Key Escrow Problem in Identity-based Encryption From Pairing"** is being submitted by **Mr Mahender Kumar,** to **School of Computer and Systems Sciences, Jawaharlal Nehru University, New Delhi-110067, India,** in the partial fulfilment of the requirements for the award of the degree of **Master of Technology** in **Computer Science and Technology**. This work has been carried out by him in the School of Computer and Systems Sciences under the supervision of Prof. **C. P. Katti**. The matter personified in the dissertation has not been submitted for the award of any other degree or diploma.

Prof. C. P. Katti
(Supervisor)

Dean, SC&SS
Jawaharlal Nehru University
New Delhi, India

# SCHOOL OF COMPUTER & SYSTEMS SCIENCES
## JAWAHARLAL NEHRU UNIVERSITY
## NEW DELHI, 110067 (INDIA)

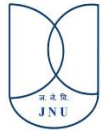

# DECLARATION

This is to certify that the dissertation work entitled **"An Approach to Remove Key Escrow Problem in Identity-based Encryption From Pairing"** in partial fulfilment of the requirements for the degree of **"Master of Technology** in **Computer Science and Technology"** and submitted to the School of Computer & Systems Sciences, Jawaharlal Nehru University, New Delhi-110067, India, is the authentic record of my work carried out during the time of Master of Technology under the supervision of **Prof. C. P. Katti**. This dissertation comprises only my original work.

The matter personified in the dissertation has not been submitted for the award of any other degree or diploma.

MAHENDER KUMAR

**(Student)**

*Dedicated to*

*My*

*Beloved Late Grandfather….*

# ACKNOWLEDGEMENT

I could never finish my dissertation without the guidance of my research Lab members, help from friends, and support from my family and relatives.

First, I would like to express my deepest gratitude to my honourable supervisor and Dean, **Prof. C. P. Katti**, School of Computer and System Science, Jawaharlal Nehru University, New Delhi, for his excellent guidance, caring, patience, and encouragement. He has been an inspiration to a great teacher and me. I would like to be very grateful towards him for providing me with an excellent atmosphere for doing research. Completing this dissertation without his valuable support and patience would have been impossible.

Secondly, I would like to thank **Prof. P.C. Saxena (Emeritus)**, School of Computer and System Science, Jawaharlal Nehru University, New Delhi, who let me experience the research of freshwater mussels in the field and practical issues beyond the textbooks, patiently corrected my writing and motivationally supported my research.

Thirdly, I would like to thank my lab mates for their fruitful support and encouragement thought-out my M. Tech dissertation. Thanks to Mr J. K. Verma, R. A. Haidri, Mr Ashok Kumar, Mr Vineet Anand, Mr Pankaj Kumar and Mr Kunal Bhaskar for providing me with their indiscipline suggestion and motivation with an excellent atmosphere for doing research.

Additionally, I would like to thank my friends and colleague for their support and care that helped me overcome setbacks and stay focused on my graduate study. Special thanks to **Mr Kashif Nawaz, Mr Ravi Shankar Soni, Mr Bhaskar Prasad, Mr Mayank Gupta, Ms Neha Anand, Mr Tarun Kumar Gupta, Mr Sunil Kumar, Mr**

**Sudhakar, Ms Ruby Rani, Ms Monika Yadav** and all friends. I greatly value their friendship and deeply appreciate their belief in me. I am also grateful to **Mr Asif Ali**, who is always willing to help, cheering me up and standing by me through the good and wrong times.

I shall ever remain obliged to the faculty members and staff of SC &SS, Jawaharlal Nehru University, New Delhi, for their cooperation and the kindness extended to me during the completion of this dissertation work. I would like to thank **Robin canteen and their Staff** for their "*Tea, Pakodas and Lunch*", where most of my thesis work was discussed with my friends. I am very thankful for their good-serving nature.

I sincerely oblige my Grandfather **Late. Sh. Nathi Singh**, whose Spiritual-Talk, Patience, Simplicity, Consistency, Encouragement, Pleasing, Bliss and Love have been continuously reflected in me through my family, I would also like to thank my **Parents, elder Brother and Sister-In-Law, two elder Sisters, two Brother-In-Law, and Nephew** and **Nieces**. They always supported, motivated, and encouraged me with their best wishes and Love.

*Mahender Kumar*...

# ABSTRACT

Key escrow refers to storing a copy of a cryptographic key with a trusted third party, typically a government agency or some other organization. Key escrow aims to ensure that law enforcement agencies can access encrypted data when necessary, for example, in criminal investigations or national security matters. However, key escrow also raises concerns about privacy and security. If the trusted third party is compromised, the stored keys could be exposed, and unauthorized parties could access sensitive information. This could result in a significant breach of privacy and potentially harm national security. In identity-based cryptography, the key escrow problem arises because a trusted third party, called the Private Key Generator (PKG), generates the private keys for all users. This means that the PKG has complete control over the private keys, which raises concerns about the security and privacy of the users. To balance security and privacy concerns, some approaches have been proposed to address the key escrow problem.

We propose an efficient democratic identity-based encryption model that balances the government's and users' rights while ensuring their security and privacy. The key objective of the proposed scheme is to provide the government with authority to monitor unlawful messages while ensuring the user's privacy for their lawful messages. To achieve this, the scheme involves two entities: PKG and PKPO. The user's partial key is escrowed at PKG, while the partial key is stored at PKPO. The latter provides a privacy service to the user by confusing their signature, which the user with their personal information can only unlock. We prove the scheme's security against IND-CCA of Type I and Type II adversary attacks in Theorem 1. We assume that two entities do not collude with each other; otherwise, the judge will apprehend the malicious entity in response to a legal complaint from the genuine entity. Additionally, we demonstrate that the proposed model is efficient for low-consumption devices since it incurs less computation cost on the client side, and overloads are shifted to the server side (PKG) on the cloud. Thus, the scheme is environment-friendly, practically applicable, and readily available.

In conclusion, the proposed scheme is ideal for wireless networks and email applications. Furthermore, we discuss the future scope of the thesis in the signature scheme and authenticated key agreement protocol.

# CONTENTS

# LIST OF TABLES

# LIST OF FIGURES

# LIST OF SYMBOLS AND NOTATIONS

| | |
|---|---|
| $M$ | Message |
| C | Ciphertext |
| $Z_n$ | Set of integers Mod n |
| (G , +) | Algebraic group with respect to the set G and the binary operation |
| (G , +, ×) | Cyclic Group with respective to the set G and two binary operation |
| G | Group (G , +) |
| H | Subgroup of Group G |
| R | Ring (R , +, ×) |
| F | Field (F , +, ×) |
| +, and × | Binary operations |
| $GF(p)$ | Finite field of order prime $p$ |
| $H(w)$ | Hash function with bit sequence $w$ as input |
| $Sign(M)$ | Signature on message $M$ |
| $e: G_1 \times G_2 \rightarrow G_T$ | Bilinear map from the group $G_1$ and $G_2$ and $G_T$ |
| x $\oplus$ y | Bitwise XOR operation between string $x$ and $y$ |
| $\perp$ | Invalid output |
| $Enc_k(M)$ | Symmetric encryption of the message $M$ under session key $k$ |
| $Dec_k(C)$ | Symmetric decryption of the ciphertext $C$ under session key $k$ |
| $x\|\|y$ | Concatenation of bit strings x and y |
| $\{0,1\}^n$ | Binary bit sequence of length n |
| $\{0,1\}^*$ | Binary bit sequence of variable length |
| $usk_{Pr}$ | User's generated private key |
| $usk_{Pub}$ | User's generated public key |
| $pkg_{Pr}$ | PKG's generated private key |
| $pkg_{Pub}$ | PKG's generated public key |
| $pkPo_{Pr}$ | PKPO's generated private key |
| $pkpo_{Pub}$ | PKPO's generated public key |
| $D_{ID}$ | Decryption key for the user with identity ID. |
| $k$ | Security parameter |
| $params$ | Publicly published parameters |
| adv | Advantage for the adversary to win the game. |
| $\epsilon$ | Negligible function |
| $Cert_{ID}$ | Certificate issued by PKG to the user whose identity ID |
| CA | Certification Authority |
| $Adv_I$ | Type I IND-CPA adversary |
| $Adv_{II}$ | Type II IND-CPA adversary |

# LIST OF ABBREVATIONS

| | |
|---|---|
| ID | User's Identity |
| IBE | Identity based encryption |
| IBS | Identity based encryption |
| PKG | private Key Generation |
| HIBE | Hierarchical IBE |
| DLP | Discrete Logarithm problem |
| DDHP | Decision Diffie-Hellman problem |
| BDHP | Bilinear Diffie-Hellman problem |
| CDHP | Computational Diffie-Hellman problem |
| GDHP | Gap Diffie-Hellman problem |
| PKI | Public key infrastructures |
| IND-CPA | Indistinguisablity chosen ciphertext attack |
| IND-CCA | Indistinguisablity chosen palintext attack |
| ECC | Elliptic Curve Cryptography |
| RSA | Rivest Shamir Adleman Algorithm |
| DKG | Distributed key Generation |
| SSN | Social Security Number |
| ANO-IBE | Anonymous Identity-Based Encryption |
| ANO-CCA | Anonymous chosen ciphertext attack |
| CL-PKC | Certificate-less Public Key Cryptosystem |
| CB-PKC | Certificate-based Public Key Cryptosystem |
| SCS | Self Certified Scheme |
| V-IBE | Variant of Identity based encryption |
| M-IBE | Modified Identity based encryption free from key escrow problem |
| PKPO | Private Key Privacy organization |
| KeyGen | Key Generation |
| ROM | Random Oracle Model |
| BasicM-IBE | Proposed Basic model of Identity-based encryption free from key escrow problems, secure against IND-CPA attack |
| FullM-IBE | Proposed Full model of Identity-based encryption free from Key Escrow Problems secure against IND-CCA attack. |

# Introduction

One of the significant challenges in cryptography is the key distribution over a public network. It is essential to securely distribute keys to ensure the confidentiality, integrity, and authenticity of messages transmitted over a network [4, 16, 35]. If keys are compromised or intercepted by attackers, the entire communication system can be compromised. This is why various encryption schemes have been developed to securely distribute keys, including identity-based cryptosystem (IBC), which uses an individual's identity, such as an email address, as a public key. IBC is a promising approach to public-key cryptography that simplifies key management and distribution [7, 18, 21, 23, 24, 29, 41, 43]. However, its adoption is still limited due to the key escrow problem and the need for further research to address its security and privacy implications.

At the same time, the key escrow problem is also a concern in cryptography that arises when a trusted third party holds a copy of a user's private key. This can occur when a government or other entity requires backdoor access to encrypted data or communication for law enforcement or national security purposes. The key escrow problem is controversial because it poses a significant risk to the privacy and security of individuals and organizations who rely on encryption to protect sensitive information. If the trusted third party is compromised or forced to disclose the private key, unauthorized parties can access the encrypted data or communication. In response to the key escrow problem, some cryptographic schemes have been developed to allow access to encrypted data or communication without requiring backdoor access to the private key. These schemes include threshold cryptography[22], where the private key is split among multiple parties, and certificate-less cryptography [31, 39], where the user generates their own public and private keys without the involvement of a trusted third-party Certificate Authority. Despite efforts to address the key escrow problem, it remains a controversial issue in cryptography and information security. There is an ongoing debate about the trade-offs between privacy and national security and the role trusted third parties should play in securing encrypted data and communication.

## 1.1     Problem Statements

In modern cryptographic algorithms [4, 5], the user generates his key pair using public and private keys. The public key is publically available to everyone in the network and is used to encrypt the message. Alternate to the public key, the private key keeps as a secret to him and is used to decrypt the message. The PKI manages each user's public key along with his identity.  No one other than the user can ever decrypt the message. So, the user has guaranteed to obtain his secret information in communication. At the same time, the user can also use encryption to encrypt criminal activity. So we can say that there are two main problems with public key encryption. The first one requires PKI to manage the certification certifying that the public key is the claimed user with an identity ID. And the second is a malicious user can encrypt criminal information.

An ID-based encryption scheme [5, 7, 18] tackles the first problem to solve the certificate management problem in PKC. Instead of using the public key, it uses the user's identity as the public key. Therefore, there is no need for a certificate to certify that the public key is the real public key of the user with identity ID.  A trusted third authority called Private Key Generator generates the private key corresponding to the identity ID; one copy is sent to the user, and the other is stored in its storage. In future, PKG may decrypt the doubtful message encrypted by the user over public communication. Thus, the second issue with PKC can be solved by monitoring the PKG's suspicious message with the help of a copy of the user's private key stored in its storage.

Malicious PKG may also decrypt the encrypted message as PKG generates the user's private key. Thus, with advantages over the public key encryption system, ID-based encryption suffers from two issues: 1) key escrow problem and 2) secure key issuing between the PKG and the user. These issues motivate the need for an efficient model to avoid such problems. Several approaches have been proposed to remove the key escrow problem [22, 31, 39, 44, 45, 46, 47], either by secure key issuing or user-chosen information. Nevertheless, the advantages come with some drawbacks. Each approach comes with new disadvantages. However, every scheme solved the key escrow problem differently. But it has become a new issue in the crypto world. These schemes discussed above show that each scheme gives full control over its private key. Indeed, giving full control over the private key to the user is also a disadvantage. User Privacy has induced two new penalties; PKG or government has no control over the user's private key, and they take no action against the user's unlawful message. Today, one hot topic in cryptography is balancing control of the Private Key for both the user and the PKG. Thus, the user has a right to privacy on their lawful message, and Government has the right to monitor the unlawful message of the user.

## 1.2    Recent Solutions

In the previous section, we have seen the limitations of PKE and IBE. Here, we discussed the existing solution of the key escrow problem in IBE and the existing solution of PKC.

### 1.2.1    Existing Solutions to key escrow Problem

Several kinds of researchers on ID-based encryption schemes have proposed avoiding key escrow problems. Boneh-Franklin [7] uses the technique of threshold cryptography [50] to distribute the master key to multiple PKG instead of one, discussed in chapter 4. Due to the massive infrastructure for managing multiple PKG, this scheme did not work so efficiently. At the same time, the HIBE scheme [9, 12] attempts to solve the issue, but the problem remains the same extensive infrastructure to manage the multiple PKG. In 2003, Gentry [45] introduced certificate-based Public Key encryption, but it needs a certificate authority to certify the user's identity and manage those certificates. To tackle this issue, certificate-less public key encryption [46] was introduced, which provides implicit authentication to the public key with user-chosen information. A new variant of IBE [31] was proposed, which uses a combination of the key issue by PKG and some information chosen by the user as a private key.

However, the most unattractive property of all solutions is each scheme is proposed to tackle the problem of key escrow with different techniques. Each scheme is supposed to have an advantage over others. By removing the key escrow problem in the existing scheme, PKG has no control over the private key in each scheme. Thus, the user may get the chance to use it privately in some criminal activity since there is no authority to monitor the user's communication.

### 1.2.2 Existing solution to key escrow

In 1993, the U. S. government declared an escrow encryption standard [51]. This scheme is based on the tamper-resistant hardware encryption device called the Clipper chip. This chip has two properties [64]:

1. SKIPJACK algorithm provides secret encryption.
2. Provide a "backdoor" for law enforcement to monitor unlawful commutation.

Since then, key escrow has been less attractive because the issue with this scheme is how to balance these two properties in a single approach. As we saw in ID-based encryption, the user's private key completely depends on the trusted third party. In 1995, Shamir [63] indicated:

*"Nowadays, even if the escrowed agent is reliable, In future, other dishonest agencies may replace it; these dishonest agencies will likely decrypt escrowed key of all users suddenly and monitor user's communication for their stake."*

Many approaches explore the problem. Shamir [40] introduced partial key escrow approaches, Micali and Ney [53] put forward a shared random function and key escrow scheme, and another improved scheme [32] which is more advantageous than the previous one.

## 1.3 Motivation

To remove the key escrow problem from ID-based encryption, several schemes have been discussed in Chapter 3. HIBE [9, 12] and threshold key issuing [7] need extra storage and computation time infrastructure. So it consumes a lot of machine cycles and is slower than other existing schemes. Certificate-based encryption [45] has the disadvantage of key revocation of the certificate. Thus, it requires a large amount of storage for the certificate and computational time for verifying those certificates. Therefore, this scheme lost the advantage of identity-based encryption. Like certificate-based encryption, Certificate-less encryption [46] uses user-chosen information but only has implicit authentication with the public key. The sender will never ensure the receiver's public key is original until communication succeeds. In contrast to an existing scheme, VIBE [31] used the user-chosen secret information and combined his confidential information and partial private generated by PKG to encrypt the message. As a result, the encryption algorithm becomes more complex. Thus, there will be a need to construct identity-based encryption, a partial key escrow problem, and monitor unlawful communication.

## 1.4    Objective of thesis

The primary objective of this thesis is to develop an efficient democratic model for identity-based cryptosystems that strikes a balance between government monitoring and user privacy. Specifically, we propose a straightforward model for transforming traditional identity-based encryption into a democratic scheme [62] that preserves the rights of both the government and the people.

1. **Government rights to monitor**: In the democratic model, the government can monitor communications through a "backdoor" mechanism, which enables authorized agencies to intercept suspicious messages and detect criminal activities. At the same time, the user's right to privacy is protected, ensuring that their personal information remains confidential and secure.

2. **User's right to privacy**: To provide the right to the user that they cannot compromise their privacy.

## 1.5    Structure of thesis

The structure of this thesis is organized as follows:

**Chapter 2** gives the mathematical background required to comprehend the proposed cryptographic model. It begins by defining fundamental mathematical concepts such as modular arithmetic, algebraic groups, finite fields, and number-theoretic assumptions. The chapter further elaborates on the concept of elliptic curves and their significance in cryptography. The chapter also introduces the notion of cryptographic hash functions and random oracles, which are essential for understanding one-way hash functions. Additionally, bilinear maps are presented as a response to different variants of the Diffie-Hellman assumption. The concept of bilinear maps enables the derivation of critical cryptographic problems such as the Bilinear Diffie-Hellman problem (BDH), Computational Diffie-Hellman problem (CDH), and Gap Diffie-Hellman problem (GDH), which ensure the security of the proposed constructions.

**Chapter 3** of the dissertation focuses on the building blocks of cryptography and their relevance to the proposed solution. The chapter begins with a review of traditional Public Key Infrastructures (PKI) and their limitations. The following section elaborates on the Identity-Based Encryption (IBE) concept and its advantages over PKI. Subsequently, the chapter compares IBE with PKI, highlighting the security aspects of IBE, including IND-CPA, IND-CCA, and ANO-IBE. The chapter also explores related work in the field of IBE, discussing various schemes that have been proposed and their relative strengths and weaknesses.

**Chapter 4** of the dissertation presents the design of a novel ID-based encryption scheme free from key escrow problems. The chapter begins by defining a model that describes the solution to avoid key escrow. This model identifies the current security threats, potential adversaries, and realistic assumptions about their capabilities. Subsequently, the chapter defines various cryptographic design goals to address the security threats identified earlier. These design goals serve as a guideline to construct a practical algorithm based on the cryptographic building blocks introduced in Chapter 3. Towards the end of the chapter, the proposed model is explored to implement various cryptographic protocols, including ID-based encryption, ID-based signature, and authenticated ID-based key exchange protocols. The chapter provides a detailed analysis of the proposed model's strengths and weaknesses and compares it with existing schemes in the literature. By presenting a novel ID-based encryption scheme that is free from key escrow problems, the chapter contributes to the advancement of secure and efficient cryptographic protocols. The proposed design goals and practical algorithm provide a framework for implementing practical solutions that address the security threats associated with key escrow.

**Chapter 5** of the dissertation provides a detailed analysis of the security and performance of the proposed ID-based encryption, ID-based signature, and authenticated key agreement protocols. The chapter begins by presenting formal proof of the security of the proposed model, demonstrating that it satisfies the required security properties, such as confidentiality, integrity, and authenticity. Subsequently, the chapter analyzes the performance of the proposed protocols in terms of point addition, exponentiation, and scalar multiplication, comparing them with existing protocols in the literature. This analysis provides insights into the efficiency of the proposed protocols and demonstrates their practicality for real-world applications.

**Chapter 6** concludes this thesis with a summary of earlier research results and our current solution's limitations. Finally, we conclude the thesis by highlighting that might be the subject of future work.

Chapter

# Mathematical Backgrounds

This chapter briefly covers the mathematical background to understand cryptographic algorithms presented in the later section. This chapter represents the fundamental of cryptography concepts.

Note that this chapter only covers the cryptographic fundamentals required to understand the remainder of the thesis. Definitions and theorems are always provided without t proof. For a more in-depth discussion on algebraic topics in this chapter, the reader is referred to [14]. More information on elliptic curves, Diffie-Hellman assumptions, and pairing-based cryptography can be found in [1].

## 2.1    Mathematics of Cryptography

In this chapter, we will be ready to understand the mathematics description by discussing the various mathematical tools and properties of cryptography. Some useful functions like field, ring, group, bilinear pairing, and the elliptic curve will be discussed here.

### 2.1.1    Modular Arithmetic

For any given positive integer $q$ and any nonnegative integer $x$, if we divide $p$ by $n$, then we get an integer quotient $q$ and an integer remainder $r$ that satisfies the following equation:

$$x = qn + r; \qquad 0 \leq r \leq n; \qquad q = \frac{x}{n}$$

Where $n$ is the largest integer less than or equal to $n$, the remainder $r$ is also known as residue or $x$ mod $n$. The integer $n$ is known as the modulus (Mod). Two

@13/0MT/016SC&SS/JNUDelhi

8

integers $x$ and $y$ are said to be congruent Mod $n$; if $x(Mod n) = y(Mod n)$, which can be written as $x = y(Mod n)$.

Properties of modular arithmetic

1. $[x(Mod n) + y(Mod n)]Mod n = (x + y)Mod n$
2. $[x(Mod n) - y(Mod n)]Mod n = (x - y)Mod n$
3. $[x(Mod n) * y(Mod n)]Mod n = (x * y)Mod n$

**Set of Residues ($Z_n$):** In modular arithmetic, the residue of an integer $a$ modulo $n$ is the remainder obtained when $a$ is divided by $n$. For example, the residue of 17 modulo 5 is 2, because 17 divided by 5 leaves a remainder of 2.

**Additive Inverse**: Suppose $x$ and $y$ are two numbers in $Z_n$, it is called the additive inverse of one another if $x + y = 0(Mod n)$. For example: In $Z_{13}$, 6 is the additive inverse of 7.

**Multiplicative Inverse**: Two numbers $x$ and $y$ are multiplicative inverse to each other if $x * y = 1(Mod n)$; for example, in $Z_{13}$, the multiplicative inverse of 6 is 11. The integer $x$ in $Z_n$ has a multiplicative inverse that exists only if $\gcd(x, n) = 1$.

Some more sets:

- $Z_n^*$: It is the subset of $Z_n$, and contains only those integers for which multiplicative inverse exists. In $Z_n$, each member contains additive inverse, but only a few contain multiplicative inverse. For example, $Z_6 = \{0,1,2,3,4,5\}$ and $Z_6^* = \{1,5\}$.

- $Z_p$: It is similar to $Z_p$ where $n$ is the prime number $p$. $Z_p$ contains all integers between 0 to $p - 1$. Each element that belongs to $Z_p$ has an additive inverse, and all components have a multiplicative inverse excluding 0. For example: $Z_{13} = \{0,1,2,3,4,5,6,7,8,9,10,11,12\}$.

- $Z_p^*$: It is similar to $Z_n^*$, where n is the prime number p and the subset of $Z_p$. In $Z_p$, only some elements have a multiplicative inverse, but in $Z_p^*$ all members have multiplicative inverse excluding 0. $Z_p^*$ contain all integers from 1 to $p - 1$. For example $Z_{13}^* = \{1,2,3,4,5,6,7,8,9,10,11,12\}$.

### 2.1.2 Mathematics of Symmetric Key Cryptography

The requirements of cryptography are the sets of integers and different operations performed on those sets. The operations applied to the elements of the combination of the different sets are called an algebraic structure. This section will briefly discuss three common algebraic structures: groups, rings, and fields.

- **Group**: A group is the set of an element that contains the operation of binary $+$ and satisfies four functions. The properties are as follows:

  1. Closure: If $x$ and $y$ are the elements of $G$, then $z = x + y$ is also the element of $G$, meaning if we apply any operation to any group element, the result will also belong to a group $G$.

  2. Associativity: If $x, y$ and $z$ are the elements of the group $G$, then
$$(x + y) + z = x + (y + z)$$

  3. Existence of identity element: For all x in a group, $G$ there exists an identity element '$e$' such that
$$e + x = x + e = x$$

  4. Existence of inverse: For each x in a group, $G$ there exists an element '$e$' such that
$$x + x^{-1} = x^{-1} + x = e$$

An Abelian group consists of a set of elements and a binary operation that satisfies the commutative, associative, and distributive properties.

  5. Commutativity: If $x$ and $y$ belong to a group, $G$ then $x + y = y + x$.

- **Finite Group**: A group, $G$ is called finite if it contains a finite number of elements; otherwise, it is called an infinite group.

- **Order of Group**: The number of unique elements present in the group is known as the group's order. If the number of an element is finite, then it is called finite order; otherwise infinite order.

- **Subgroup**: A subset $H$ is called the subgroup of a group $G$ if $H$ it is a group concerning the operation on $G$. In other words, if $G = <x, +>$ is a group and $H = <y, +>$ is a group under the same operation, and $y$ is a non-empty subset of $x$, then $H$ is called a subgroup of $G$. This definition yields:

1. If $x$ and $y$ are the members of both groups, then $z = x + y$ is also the member of both groups.
2. The same identity element exists for both.
3. If $x$ belongs to both groups, then the inverse of $x$ also belongs to both groups.
4. Each group is itself a subgroup.

- **Cyclic Subgroup**: A subgroup is said to be cyclic if the power of an element of the group can generate it.

$$x^n = x + x + \cdots \ldots + x$$

- **Cyclic Group**: It is the group that contains its own Cyclic Subgroup. The Component that can generate a cyclic subgroup can also produce the whole group itself; that component is called the generator of the group. If $g$ is a generator, then the elements in the finite cyclic group can be written as

$$\{g^0, g^1, \ldots g^{n-1}\}, \quad where \ g^n = e$$

Note: A cyclic group can have many generators.

Example: The group $G = < Z^*, +, \times >$ is a cyclic group with two generators, $g = 3$ and $g = 7$.

- **Ring**: A ring $R = < \{..\}, +, \times >$ is an algebraic structure with two operations. The first operation $(+)$ satisfies all five properties of the abelian group: Closure, Associativity, Commutativity, Existence of inverse and Existence of identity. The second operation $(\times)$ is distributed over the first operation and satisfies only the two properties: Closure and Associativity. The ring in which the second operation satisfies the commutative property is called a commutative ring.

- **Field**: A field $F = < \{..\}, +, \times >$ is a type of commutative ring in which the second operation satisfies all five properties defined for the first operation, except that the identity of the first operation has no inverse.

- **Finite Fields**: A field with a finite number of an element is called a finite field. Galois demonstrated that the field has a finite number of the component that must be $p^k$, where $p$ is a prime number and $k$ area is a positive integer. A Galois field, $GF(p^n)$, is a finite field with $p^n$ elements. While $n = 1$, we have $GF(p)$ field.

## 2.2    Elliptic-Curve Cryptosystem

The properties and functions of elliptic curves have been studied in mathematics for 150 years. In 1985 Victor Miller [56] and Neal Koblitz [55] suggested an elliptic curve as a mathematical tool in cryptography known as the elliptic–curve cryptosystem.

### 2.2.1    Definition of Elliptic Curve

An Elliptic-curve over a finite field is a smooth non-singular cubic projective curve of genus one defined over $k$ with distinguished $k$ rational points. By, non-singular means all three roots of EC must be distinct. Over any field $F$, an irreducible projective curve is a compact manifold that is topological as a sphere with handles.

### 2.2.2    The general form of the Elliptic Curve

The following equation can define any elliptic curve:

$$y^2 = x^3 + ax + b$$

Where $x$ is not a continuous point that is chosen from a particular field $GF(P)$ or $GF(2^k)$, figure 2.1 shows the elliptic curve [8] of the equation $y^2 = x^3 - x + 1$.

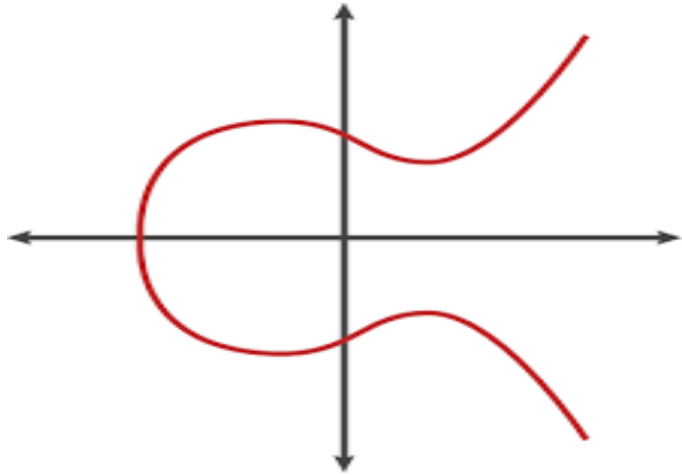

Figure 2.1 Graphical representation of the elliptic curve $\boldsymbol{y^2 = x^3 - x + 1}$

Let $E$ be an elliptic curve over $F$ defined by the Weierstrass equation as follow:

$$y^2 + a_1xy + a_3y = x^3 + a_2x^2 + a_4x + a_5$$

If P is a rational point on elliptic curve $E$ and $l$ is a line through $P$ with a rational slope, it is not necessarily true that $l$ intersects $E$ in another rational point. However, if $P$ and $Q$ are two rational points on the elliptic curve $E$, then the line $PQ$ intersects $E$ in a third rational point $R$. This permits us to generate many new rational points from old ones. And also permits us to define a group operation on $E(k)$ for any elliptic curve defined over any field $k$.

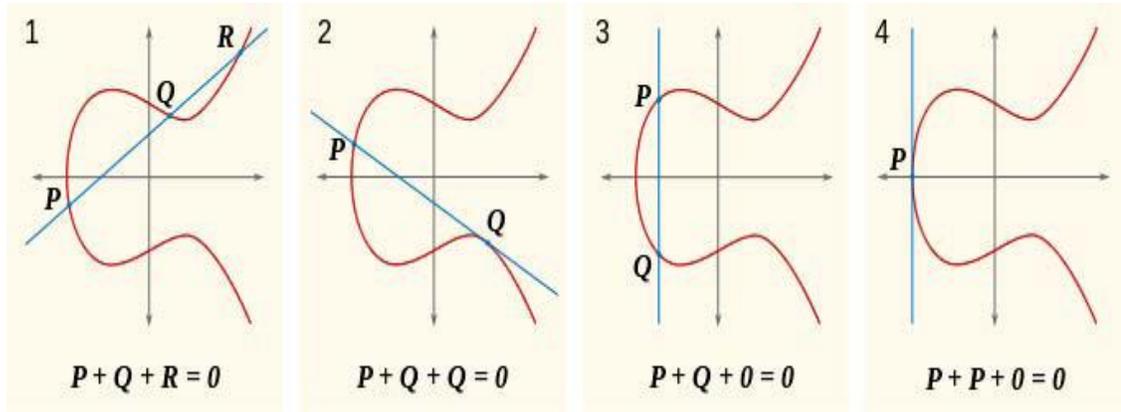

Figure 2.2  Rational points on line over an elliptic curve.

Figure 2.2 denote that the sum of three rational points on line $l$ over elliptic curve $E$ should be zero [17]. Here, zero is a point at infinity. The evaluation of $P + Q = R$ is purely algebraic. The coordinates of R are rational functions of the coordinates of $P$ and $Q$, and can be computed over any field. By adding a point to itself repeatedly, we can compute $nP = P + P + \cdots + P$ for any positive $n$. We also define $0P = 0$ and $(-n)P = -nP$. Thus, we can perform scalar multiplication by any integer $n$.

### 2.2.3  Why Elliptic Curve Cryptography?

Elliptic curve cryptography (ECC) is a modern public-key technique based on elliptic curve theory. It has gained popularity recently due to its robustness, speed, and efficiency. One of the main advantages of using ECC is the ability to create smaller and faster keys than traditional methods, such as RSA. ECC generates a pair of keys using an elliptic curve equation, a mathematical formula describing a curve on a plane. The keys consist of public and private keys used for encryption and decryption, respectively. The security of the keys is based on the difficulty of solving a mathematical problem related to the elliptic curve equation.

Compared to RSA, which requires a 1024-bit key for a certain level of security, ECC can provide an equivalent level with only a 164-bit key. ECC is much more efficient in computation power and battery resource usage, making it particularly well-suited for mobile applications. Moreover, the key size of ECC doubles every ten years, which means that traditional methods cannot be used due to the large key size required for maintaining security. ECC offers a practical solution to this problem by allowing for smaller key sizes without compromising security.

Table 2.1 RSA key length of Some Organisation

| Organization | RSA key length(in bits) |
|---|---|
| ICICI Bank | 1024 |
| Amazon | 2048 |
| eBay | 2048 |
| Online SBI | 2048 |
| Facebook | 1024 |
| Canara Bank | 2048 |

Table 2.1 shows some currently used RSA key lengths by some organizations. If the key size increase, it increases security, but it causes a severe problem. Decryption will be eight times slower if we double the RSA key length. Table 2.1 gives the RSA key length of some organizations, and Table 2.2 gives the security level of ECC and RSA. Ciphertext size also becomes large. The encryption speed is also infected with a large key length, which is slower by a factor of 4. Table 3.2 gives the security level of the ECC and RSA scheme. Tables 2.1 and 2.2 clearly show that ECC takes less key length than RSAit is more efficient.

Table 2.2 RSA and ECC key Sizes [65]

| Key Size (in bits) | 80 | 112 | 120 | 128 | 256 |
|---|---|---|---|---|---|
| ECC | 160 | 185 | 237 | 256 | 512 |
| RSA | 1024 | 2048 | 2560 | 3072 | 15360 |

**Application of ECC**: ECC has a wide range of applications in various fields, from secure communication to cryptocurrency and IoT devices. Its efficiency, scalability, and strong security guarantees make it an ideal choice for modern cryptography applications.

## 2.3   Cryptographic Hash Function

A cryptographic hash function [57] is a function that takes a variable-length string as an input and gives a Fixed-length string, i.e. message digest. $H: \{0,1\}^* \rightarrow \{0,1\}^k$ where, $k$ is the length of a message digest. Let's take a function $f(x) = y$ that maps $x$ to the image $y$. $x$ is called a pre-image of $y$. The output is called hash value or message digests. Here we use $y = H(x)$, which denotes applying the hash function into variable-length message $x$ and gives fixed length digest $y$. The hash function should follow some characteristics:

1.  $x$ should be variable length, and $y$ is fixed length.

2.  For given $x$, it's easy to compute $y$, but vice versa should be very tough, which means the hash function should be the one-way function.

3.  Two messages don't have the same message digest.

4.  The hash function must be easy to compute.

Suppose x and y are messages; then $H(x) = H(y)$ is infeasible. Today, the Hash function is used in various cryptographic techniques like message authentication code (MAC), digital signature, Random sequence generators used in key agreements, authentication protocol, etc. Hash function needs to satisfy the four main properties:

1.  **Pre-image Resistance**: Given a digest $y = H(x)$, it is computationally infeasible to compute x. That is, the computational cost of getting the input $x$ must be $\geq 2_k$, where $H(x) = y$ and $|y| = k$.

2.  Pre-image resistance is the hash function for which pre-image can't be solved efficiently.

3.  **Second pre-image resistance**: Given message $x$, it is computationally infeasible to compute a different message $x_0$ with the same message digest, i.e. $H(x) = H(x_0)$ is infeasible. It is called second pre-image resistance.

4.  **Collision resistance**: Finding two messages with the same message digest is impossible. That means if $x$ and $x_0$ are two different messages, then $H(x) = H(x_0)$ is incomprehensible. This property is known as collision resistance.

Hash functions are helpful in a wide range of practical applications. For example, hash functions act as one-way functions in password databases to lighten the sensitivity of the stored content. Besides, hash functions are also valuable tools for data authentication and integrity checking.

### 2.3.1 Random Oracle model

The Random Oracle model (ROM) was introduced in 1993 by Bellare and Rogaway [58]. A Random Oracle is a theoretical black box that gives a uniformly Randomly chosen result from its output domain for each unique query. A Random Oracle is deterministic, i.e. given a particular input, it will always produce the same output.

The behaviour of this model is given as follows:

1. When any new message comes, Oracle creates a fixed size of the digest for that message and saves the message and digest in the Oracle record.

2. When any message exists, and digest exists for that message, then Oracle puts the message digest on their record.

3. The digest for any new information is independently chosen from the previous digest.

In an ideal model, hash functions can be considered random oracles. The hash function output will look like perfect random bit sequences if the hash function is ideal. Therefore, hash functions are often examined as Random Oracles in security proofs. Such security proofs are called proven secure in ROM. The next step of these security proofs is replacing the Random Oracle access with the computation of an appropriately chosen (hash) function [58]. Algorithms that are not requiring such a system in their security proof are said to be proven secure in the standard model.

### 2.3.2 Pigeonhole principle

The Pigeonhole principle is a fundamental concept in mathematics that states that if there are more objects than containers to put them in, then at least one container must contain more than one object. In other words, if n+1 or more objects are placed into n containers, then at least one container must contain more than one object. The principle is often used to prove various theorems in combinatorics, number theory, and other fields of mathematics. It can be applied in various contexts, such as counting problems, graph theory, and probability theory.

## 2.4  Pairing-Based Cryptography

The main idea of pairing-based cryptography [1, 3] is to map between two essential groups. It allows a new scheme based on reducing one problem to another, which means reducing the hard problem from one group to the problem more accessible than the first one in another group.

### 2.4.1  Bilinear Maps

Bilinear Map [6] allows mapping between different groups. Let $G_1$ be the cyclic additive group with generator p. The bilinear map is also called pairing because it allows a pair of elements from $G_1$ and $G_2$ to another group $G_T$. Suppose $G_1$, $G_2$, and $G_T$ are cyclic groups with large prime order $q$. Generally, $G_1$, and $G_2$ are the additive group, and $G_T$ is the multiplicative group. A bilinear pairing is described as $e: G_1 \times G_2 \to G_T$ that satisfy the bilinear property:

$$e(xP, yQ) = e(P, Q)^{xy} \quad for \; all \; P \in G_1, Q \in G_2 \; and \; all \; x, y \in Z_p$$

It means if $P$ is the generator of $G_1$ and $P$ is the generator of $G_2$, then $e(P, Q)$ is the generator of $G_T$. The mapping is computable if some algorithm can efficiently compute $e(P, Q)$ for $P, Q \in G_1$. If $G_1 = G_2$ pairing is called symmetric otherwise, pairing is known as asymmetric. If $G_1 = G_2 = G_T$, pairing is called a self-bilinear map.

### 2.4.2  Bilinear pairing

Suppose $G_1$ is a cyclic additive group, $G_2$ is a cyclic multiplicative group of the same order $q$, and $P$ is a generator of $Z_q$. A bilinear pairing is a map $e: G_1 \times G_2 \to G_T$ that satisfies the following properties:

1. *Bilinearity*: For every $P, Q, R \in G_1$,

$$e(P, Q + R) = e(P, Q)e(P, R)$$
$$e(P + Q, R) = e(P, R)e(Q, R)$$

for any $x, y \in Z_q$

$$e(xP, yQ) = e(P, Q)^{xy} = e(xyP, Q)$$

2. *Non-Degeneracy*: If everything maps to identity, then it is undesirable; if $P$ is a generator of $G_1$, then $e(P, P)$ is a generator of $G_T$, that means if there exists $P \in G_1$ such that $e(P, P) \neq 1$, where 1 is the identity element of $G_T$.

3. *Computability:* An algorithm must efficiently compute $e(P, Q)$ for every $P, Q \in G_1$.

Here, $G_1$ and $G_2$ are as an additive notation and $G_T$ with a multiplicative notation. In general, $G_1$ and $G_2$ are the groups of points on an elliptic curve, and $G_T$ will denote the multiplicative subgroup of the finite field. Map $e$ will be Tate pairing or Weil pairing on an elliptic curve over a finite field.

### 2.4.3 Mathematical problem

Now, we are ready to describe some mathematical problems.

1. **Discrete Logarithm (DLP) problem**: Given two random integers $P \in G_1$ and $P \in G_2$, compute an integer $x$, such that $Q = xP$, where $x \in Z_q$. The security of these algorithms relies on the assumption that the DLP is hard to solve.

2. **Bilinear Diffie-Hellman (BDH) problem**: Given $x, y, z \in Z_q$ , and $< P, xP, yP, zP >$ compute $e(P, P)^{xyz} \in G_T$. The security of these algorithms relies on the assumption that the BDH is hard to solve.

3. **Computational Diffie-Hellman (CDH) problem:** Given $x, y \in Z_q$ and $< P, xP, yP >$, compute $xyP$ . The security of these algorithms relies on the assumption that the CDH is hard to solve.

4. **Decision Diffie-Hellman (DDH) problem [11]:** Given $x, y, z \in Z_q$ , and $< P, xP, yP, zP >$ check whether $z = xy Mod q$. The security of these algorithms relies on the assumption that the DDH is hard to solve.

5. **Gap Diffie-Hellman (GDH) problem**: A group of problems where DDHP is easy while CDHP is hard.

The security of the proposed model discussed in this dissertation is based on the GDH problem.

## 2.5    Security Gaols

This section defines primary security goals and their notation applied throughout the remaining part of the dissertation.

1. **Confidentiality**: This concept ensures that only authorized individuals or entities can access sensitive information. It is essential to keep the information confidential to prevent unauthorized disclosure, misuse, or information theft.

2. **Integrity**: This concept ensures that information is not altered, tampered with, or destroyed without authorization. Integrity ensures data remains accurate, consistent, and trustworthy throughout its lifecycle.

3. **Authentication**: This concept ensures that individuals or entities claiming to be who they are have been authenticated and are authorized to access the information. Authentication can be achieved through various means, such as passwords, biometrics, or digital certificates.

4. **Authenticity**: This concept ensures that the information received is from the claimed entity and has not been tampered with or modified in transit. Authenticity is important to prevent data tampering, impersonation, or fraud.

5. **Non-repudiation**: This concept ensures that the sender of a message cannot deny sending the message once it has been sent. Non-repudiation assures that the information received is authentic and has not been modified during transit. It is essential for legal and regulatory compliance.

## 2.6    Summary

This chapter provided an overview of the basic mathematical concepts required to understand the cryptographic model described in Chapter 3. The chapter began with an introduction to modular arithmetic, algebraic structures, groups, rings, and fields, which were fundamental to cryptography. The section then covered elliptic curve cryptography, including elliptic curve definitions, graphical representations, and applications. The chapter also discussed hash functions and random oracles and the differences in security between random oracle assumptions and the standard model. It introduced various versions of the Diffie-Hellman problem, starting from the discrete logarithm assumption and leading to the Gap Diffie-Hellman assumption. Finally, the chapter concluded with an overview of basic cryptographic terminology, providing a comprehensive understanding of the cryptographic concepts and techniques necessary for the following chapters.

# Cryptographic Building Blocks

This chapter provides an overview of the essential cryptography components employed in constructing a fair model of ID-based encryption free from key escrow issues. It introduces Public Key Infrastructures (PKIs) and their associated constraints. Subsequently, the chapter delves into Identity-Based Encryption (IBE), an alternative to conventional PKIs, and examines its benefits and drawbacks. Additionally, various solutions to the key escrow problem are presented. The chapter also outlines different security definitions. Finally, Distributed Key Generation (DKG) is discussed as a possible solution to IBE's inherent key escrow challenge.

## 3.1    Public Key Cryptosystem

A class of asymmetric cryptosystems based on an algorithm takes two different keys; one for encryption and another for decryption. The user generates the key pair ($usk_{Pr}$, $usk_{Pub}$). The private key is kept secret to himself; on the other hand, the Public Key is publicly available to everyone. The pair of private and public keys allows secure communication between two never met parties. To verify the linking of a Public Key to the corresponding user, a trusted authority must prevent an impersonation attack. All Public Keys are authenticated by the infrastructure known as Public Key Infrastructure (PKI). The sender requests the CA to issue the received public key whenever needed. On receiving the request, the CA provides the receiver with a public key if only the sender's identity is valid. Now, the sender encrypts/signs the message with a public key, yields the ciphertext and sends it to the receiver over the insecure medium.

Consequently, a receiver decrypts/verifies the encrypted/ signed message with his Private Key. One of the good things about this cryptosystem is that the private key must not be transmitted or revealed to anyone. Thus, security increased. It gives the facility a digital signature so the user can never deny its existence. From a security point of view, PKI believed in trusting the user's key rather than theirs. Suppose Eve has made the PKI, who ensures that his Public Key is identified as Alice's Public Key. So, Eve can modify all of Alice's communication as he has Alice's Public Key corresponding to Alice's Private Key. Thus, the Public Key system is required depending on the infrastructure that authenticates whether key pairs belong to claimed user. In the real world, this can be achieved with the help of the certificate authority.

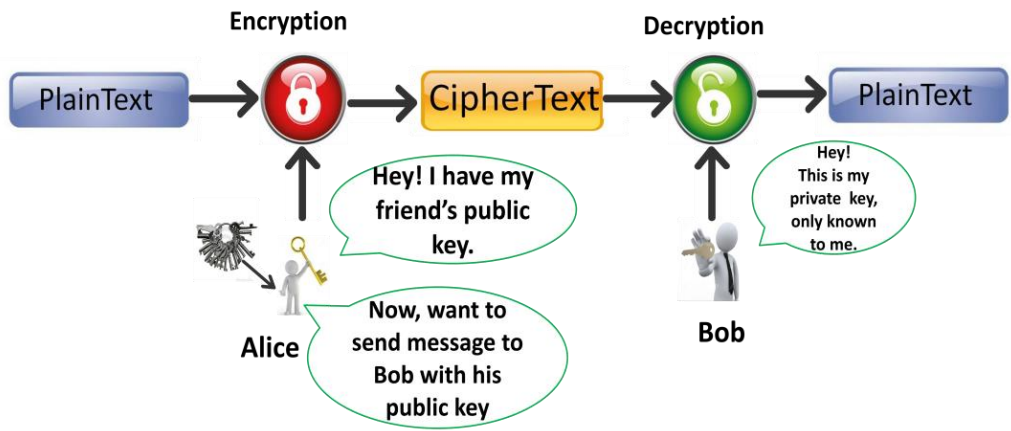

Figure 3.1 Public Key Cryptography

### 3.1.1 Certification Authority

In a Public Key Infrastructure (PKI), a Certificate Authority (CA) is a trusted entity that issues digital certificates to entities (such as individuals, organizations, or devices) to verify their identity. A digital certificate is an electronic document containing a public key, the identity of the entity to which the certificate is issued, and other information such as the CA's digital signature, which validates the certificate's authenticity [60]. A CA acts as a trusted third party that verifies the identity of entities and issues digital certificates that attest to their identity. The CA's digital signature on the certificate ensures that the certificate is authentic and has not been tampered with. When a user wants to establish a secure communication channel with another entity, they can use the digital certificate to verify that entity's identity and encrypt and authenticate their communication.

In a PKI, there can be one or more CAs that issue digital certificates. A hierarchical structure is often used in which a root CA issues certificates to intermediate CAs, which issue certificates to end entities. This structure allows for scalability and the ability to delegate certificate issuance to different organizations. A CA's role in a PKI is critical, as the trustworthiness of the entire system depends on the trustworthiness of the CA. Therefore, CAs are subject to strict security and operational requirements, including the use of secure cryptographic algorithms, regular audits, and secure storage of private keys.

## 3.2 Identity-based Cryptosystem

The concept of identity-based encryption first introduces by Adi Shamir [5], co-inventor of the RSA system, in 1984. It uses the user identity as a Public Key instead of the digital signature for encryption and signature verification. User identity can be anything they can uniquely identify, such as email-id, phone number, SSN, etc. Shamir's innovation was to eliminate the need for generating and managing the users' certificates. This feature reduces the complexity of the cryptosystem. This makes it more efficient to provide cryptography for novice users. Shamir's scheme [5] is based on the integer factorization of RSA. This scheme is built only for signature and verification. It becomes an open challenge for all researchers. Since then, many ID-based encryption schemes [22, 31, 39, 44, 45, 46, 47] have been introduced. In 2001, Boneh and Franklin [7] were the first to propose an identity-based encryption scheme based on bilinear pairing. Moreover after, Lynn [42] and cocks [34] were also two of several Identity-based encryption schemes.

### 3.2.1 Definition

This scheme consists of four algorithms (Setup, Extract, Encrypt, and Decrypt) as shown in figure 1.2 and runs as follows:

- **Setup:** On input of a security parameter $k$, outputs a master secret $msk_{Pr}$ and public parameters $params$.

- **Extract:** Takes public parameters $params$, the master secret $msk_{Pr}$, and an $ID$ as input and returns the private key $usk_{Pr}$ corresponding to $ID$.

- **Encryption:** Returns the encryption $C$ of the plaintext message $M$ on the input of the public parameters $params$, $ID$, and the arbitrary length message $M$.

- **Decryption:** Decrypts the Ciphertext $C$, back to the plaintext message $M$ on the input of the private key, $D_{ID}$ corresponding to the receiving identity ID.

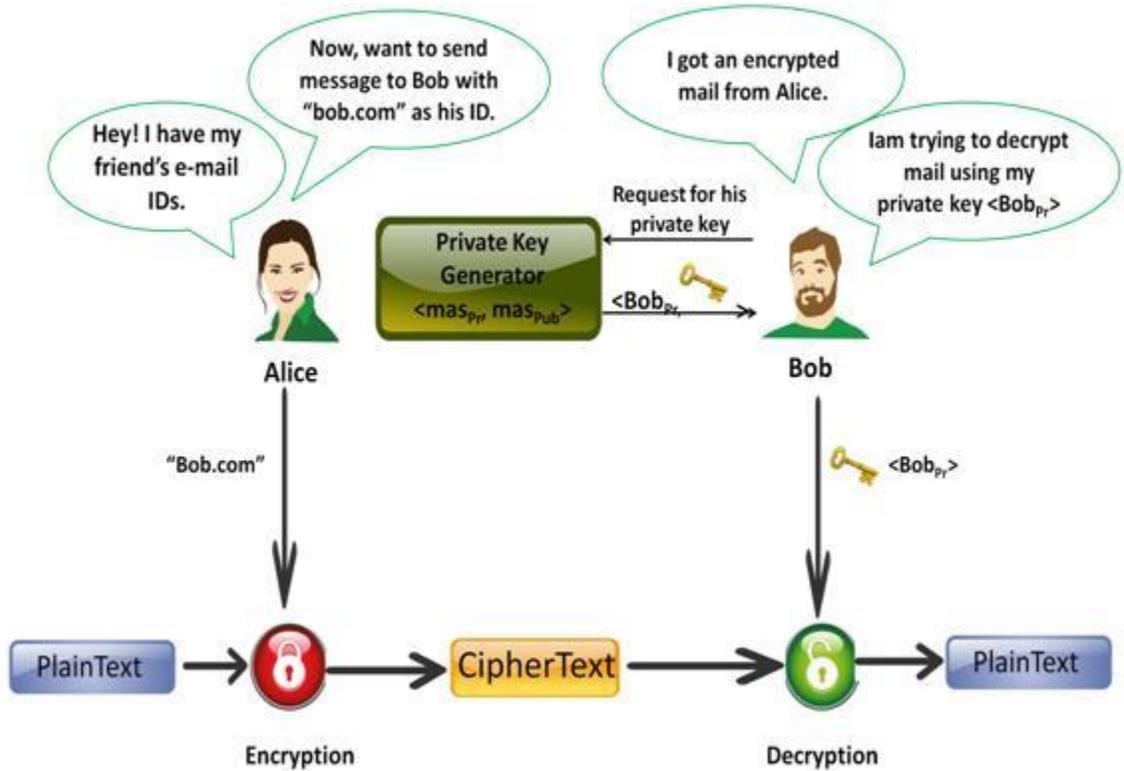

Figure 3. 2 Identity-based encryption

### 3.2.2 IBE vs PKC

Public Key Cryptography (PKC) and Identity-Based Encryption (IBE) are two cryptographic systems used for secure internet communication. The main difference between the two lies in their approach to generating and verifying key pairs [15]. PKC and IBE differ in their approach to generating and verifying key pairs. PKC relies on users developing their key pairs and verifying public keys through digital certificates. At the same time, IBE uses identity information to derive public keys and relies on a trusted authority to generate and distribute private keys.

**Advantage:** When compared to the traditional PKC, IBC has the following benefits:

- **System complexity**: PKI systems are complex infrastructure due to the support of revocation lists and hierarchical organization of CA. On the other hand, the IBE scheme has only one PKG that serves to understand fully the IBE scheme that lightens complex infrastructure requirements.

- **User Friendly**: Users with no knowledge of cryptographic primitives no longer have to make aware of the decision on the key length of their key. The Public Key in an IBE scheme is formulated to be transparent to users without knowledge of cryptography. Thus, on average, for any user, it is easy to memorize the username or e-mail address rather than authenticating the Public Key.

- **Certificate Management**: In traditional PKI, there is a large number of users who get certifications from a trusted authority. As shown in the previous section, we have seen that it is challenging for management and distribute the user's certificate. While this could be avoided in the IBE scheme with the help of PKG, which generates the user's Private Key, using their unique identification entity, on user request.

**Disadvantage:** When compared to the traditional PKC cryptosystem, the IBC has the following disadvantages:

- **Single point failure**: Every user's Private Key is generated by the PKG in the system, consequently suffering the single point of failure. A new user can no longer get their Private Key if PKG disconnects the communication due to many extraction requests.

- **Key escrow**: The PKG extracted and stored the user's private key. A malicious PKG can use this information to tap on the insecure channel between two users. The inherent property in ID-based encryption that Private Keys have to share with PKG is called key escrow [20]. On the contrary, traditional PKI only authenticates private and public keys; it does not have key escrow.

- **Public Key revocation**: The generic IBE scheme does not support the revocation of Public Keys. Although, if the recipient is hasty towards his private key's privacy, his Private Key can get compromised. Indeed, several researchers have worked on the same issue [10, 25]. It requires an extra infrastructure that makes the generic IBE system more complex. The main disadvantage of revoking of receiver key is that he can no longer receive an encrypted message. Thus, the practical solution to this problem in [7] is to append the expiration date along with the Public Key so that Public Key will have no use after the expiration date. Similarly, traditional PKI publishes revocation lists for a solution to the same issue, but this list makes PKI more complex.

### 3.2.3 Application of Identity-based encryption

Apart from Data encryption, digital signature, and key management, several other practical applications exist. Boneh and Franklin [7] show some applications:

- **Public Key Revocation**: In an IBE scheme, it is easy to make key expiration by encrypting the message using the Public Key "receiver-ID||current-time". Unlike PKC, the receiver will query the PKG to obtain the new Private Key instead of getting new certificates eve time. Thus, the IBE scheme is a very efficient and powerful way of implementing ephemeral Public Keys. This proposal can also help send messages in the future since the receiver cannot decrypt the message until he obtains the Private Key for the date specified by the sender from the PKG.

- **Managing user credential**: encrypt the message using the Public Key "receiver-ID||current-time||Clearance-level" so that the receiver can decrypt the message only if he has given Clearance. Consequently, PKG grants and revokes the user credential.

- **Delegation of the Private Key**: Suppose a manager acts as a Private Key Generator in a company and has several assistants responsible for different jobs. The manager gives his assistants the Private Key corresponding to their responsibilities. So, according to his responsibility, each assistant can decrypt the message but cannot decrypt the message to another assistant because the manager has his master key so that he can decrypt all messages.

- **Forward secure encryption**: In a forward-secure encryption scheme [49], each time the receiver's Private Key regularly evolves so that the Private Key of a particular period is compromised, every message encrypted in the past will be secure.

- **Authenticated Key Agreement**: Diffie-Hellman [4] was the first who establish the first feasible approach for constructing a shared secret over an insecure communications network, but no user authentication is there. Several systems [26, 28, 52, 48] are introduced for user authentication. Nan Li proposed that the scheme [13] provides user authentication with the help of an authentication server and the hash algorithm. M. Kumar et al. [38] introduce the ID-based authenticated key exchange protocol and remove the attack subjected to the DH scheme.

### 3.2.4 Security of IBE

Unlike the Public Key system, IBE also follows the same security aspects. Therefore, definitions of security are often discussed. In literature, the most favourable security aspects are indistinguishability under chosen plaintext attack (IND-CPA) and indistinguishability under chosen Ciphertext attack (IND-CCA).

**Indistinguishability under Chosen Plaintext Attack**

Indistinguishability under Chosen Plaintext Attack (IND-CPA) is a security notion used to evaluate the strength of encryption schemes, particularly in symmetric-key encryption. IND-CPA means that an adversary cannot distinguish between two encryptions of plaintexts of their choice, even if they have access to the encryption algorithm and can perform chosen plaintext attacks. In other words, the adversary should not be able to learn any information about the plaintext just by observing the ciphertexts.

To formally define IND-CPA, we consider a game between an adversary and a challenger. The adversary selects two plaintexts of their choice and sends them to the challenger. The challenger randomly selects one of the plaintexts, encrypts it, and sends the resulting ciphertext back to the adversary. The adversary's goal is to identify which plaintext was encrypted correctly. A secure encryption scheme is said to satisfy IND-CPA if the adversary's success probability in this game is negligibly close to 0.5, i.e., the adversary is essentially guessing at random and cannot learn any information about the plaintexts from the ciphertexts. IND-CPA is a strong notion of security for encryption schemes and is considered a minimum requirement for most practical applications. Suppose an encryption scheme is not IND-CPA secure. In that case, it is vulnerable to chosen plaintext attacks, where an adversary can manipulate the plaintexts to learn about the encryption key or other sensitive information.

More formally, an IBE system is IBE-IND-CPA secure if, for every adversary with an advantage Adv in winning the IBE-IND-CPA game illustrated in Game 1; there exists a negligible function ε such that $Adv \leq \epsilon$.

---

**Game 1:** Generic IBE-IND-CPA [2]

**Aim**: An adversary is challenged to check the IND-CPA security of an IBE scheme by a game.

**Output**: This IBE-IND-CPA Game helps to define the concept of IND-CPA security for IBE schemes.

1. The challenger runs $< msk_{Pr}, params > < -Setup(1^k)$ and returns $params$ to the adversary.

2. The adversary can start querying an Oracle $O_{extract}(ID_i)$ that returns a private key $D_{ID} < -Extract(params, msk_{pr}, ID_i)$ corresponding to an adversary-defined identity $ID_i$.

3. The adversary picks two equal-length plaintext messages $M_0$ and $M_1$ and an identity $ID_{Chal}$. The adversary honestly passes $< M_0, M_1, ID_{Chal} >$ to the challenger.

4. The challenger picks a Random bit $b$ and executes $C < -Encrypt(params, M_b, ID_{chal})$. The challenger gives C to the adversary.

5. The adversary continues querying the Oracle $O_{extract}(ID_i)$ adaptively.

6. The adversary outputs a bit $b'$ based on the Ciphertext C. If $b = b'$ adversary wins the game. If $b \neq b'$ or if the adversary queried the Oracle $O_{extract}(ID_i)$ with $ID_i = ID_{chal}$ during step 2 or step 5, the adversary loses the game.

---

*Indistinguishability Under Chosen Ciphertext Attack*

Indistinguishability under Chosen Ciphertext Attack (IND-CCA) is a stronger security notion than IND-CPA used to evaluate the strength of encryption schemes, particularly in public-key encryption. Like IND-CPA, IND-CCA ensures that an adversary cannot distinguish between two encryptions of plaintexts of their choice. However, it also considers the scenario where an adversary has access to a decryption oracle, which allows them to obtain the decryption of any ciphertext of their choice, except for the challenge ciphertext used in the security game.

To formally define IND-CCA, we again consider a game between an adversary and a challenger. The adversary is given access to a decryption oracle and can make a polynomial number of queries to this oracle. The challenger selects a random key pair and encrypts a random message, except for a challenge ciphertext. The adversary's goal is to correctly guess the bit that was used to randomly determine the encryption key, which is concealed in the challenge ciphertext. The adversary can submit ciphertexts to the decryption oracle to obtain the corresponding plaintexts.

A secure encryption scheme is said to satisfy IND-CCA if the adversary's success probability in this game is negligibly close to 0.5. This means the adversary cannot distinguish between the encryption of the two possible messages, even when given access to a decryption oracle that allows them to obtain plaintexts for any ciphertext except the challenge ciphertext.

IND-CCA is a stronger notion of security than IND-CPA and is typically required in applications where the adversary can obtain the decryption of ciphertexts of their choice. Examples of such applications include digital signatures and hybrid encryption schemes. In the non-adaptive case, steps 6 and 7 from Game 2 are prohibited. More precisely, an IBE scheme that satisfies Game 2 is considered IND-CCA2 secure.

| **Game 2:** Generic IBE-IND-CCA [2] |
|---|
| **Goal**: An adversary is challenged to check the IND-CCA security of an IBE scheme by a game. <br> **Result**: This IBE-IND-CCA Game helps to define the concept of IND-CPA security for IBE schemes. <br>     1. The challenger runs $< msk_{Pr}, params > < -Setup(1^k)$ and returns $params$ |

to the adversary.

2. The adversary can start querying an Oracle $O_{extract}(ID_i)$ that returns a private key $D_{ID} < -Extract(params, msk_{pr}, ID_i)$ corresponding to an adversary-defined identity $ID_i$.

3. The adversary can start querying another Oracle $O_{Decrypt}(D_{IDi}, C_j)$ that returns a plaintext $M_j < -Decrypt(D_{IDi}, C_j)$ corresponding to an adversary-defined Ciphertext $C_j$ and identity $ID_i$.

4. The adversary picks two equal-length plaintext messages $M_0$ and $M_1$ and an identity $ID_{Chal}$. The adversary honestly passes $< M_0, M_1, ID_{Chal} >$ to the challenger.

5. The challenger picks a Random bit $b$ and executes $C < -Encrypt(params, M_b, ID_{chal})$. The challenger gives C to the adversary.

6. The adversary continues querying the Oracle $O_{extract}(ID_i)$ adaptively.

7. The adversary continues querying the Oracle $O_{decrypt}(D_{IDi}, C_j)$ adaptively.

8. The adversary outputs a bit $b'$ based on the Ciphertext C. If $b = b'$ adversary wins the game. Otherwise, the adversary loses the game. If the adversary queried the Oracle $O_{extract}(ID_i)$ with $ID_i = ID_{chal}$ during step 2 or step 6 or if the adversary queried the Oracle $O_{decrypt}(D_{IDi}, C_j)$ with $C_j = C$ during step 3 or step 7, the adversary also loses the game.

## 3.3    Key escrow

As we have seen, the user's identity (ID) is directly used as the Public Key, and the corresponding Private Key is generated by the PKG and stored in it. Therefore, an unusual property is inherent in the proposed IBE scheme. This property is called "*key escrow.*"Moreover, the second is after the extraction of a user's Private Key, the PKG sends it over the secure channel, making the channel secure is difficult. Key escrow is a situation in which an authorized party stores the users' Private Keys. Under a certain criminal condition, the authorized party may access those keys and decrypt the private communication. These authorized parties might include government or any commercial department that may wish to tap the users' secret communication. However, what happens when these authorized parties impersonate the users by revealing their private keys to criminals? This situation is known as the key escrow problem, which may create a serious issue for users in certain situations, as shown in Figure 3.3.

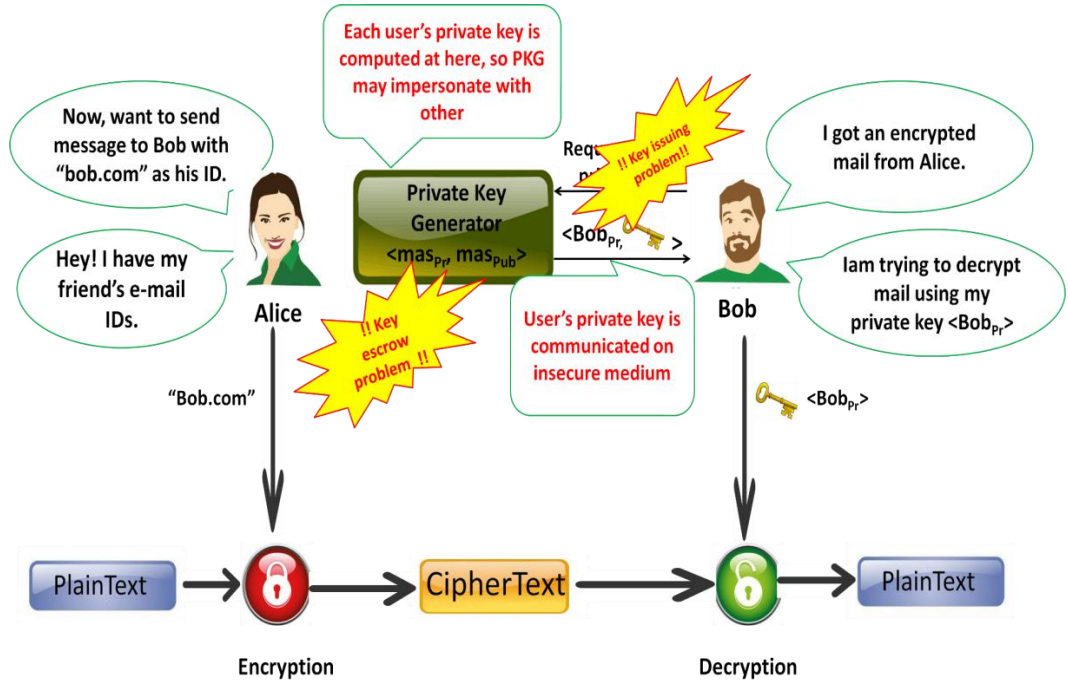

Figure 3. 3 Key Escrow Problem in Identity-based encryption

The trusted authority (PKG) is the one who never has pre-enrolled and wants to share a secret over an authenticated physical medium between all users joined to the same network. That means if two users or confidential channels. They may set up the required channel if both users trust this authority. However, what does "trust" mean here? What parameters are there which define how much possible for a user to trust the authority? The Girault proposed in the scheme [22] defines the levels of trust in PKG into three categories. As shown in Table 2.1, level 1 refers to those PKG who can easily compute a user's private key to impersonate any user without being detected, e.g. IBE. Level 2 refers to PKG, which cannot compute a user's private key but still can impersonate the user by generating fraud certificates, e.g. certificates signature scheme. That is why it requires level 3, in which PKG cannot find computer users' private keys and cannot impersonate any user without being recognized. Level 3 is the most advantageous one. The certificate-based scheme is the one which achieved Level 3.

Table 3.1 Level of trust to PKG

| Level | Can PKG compute user's secret key? | Can PKG impersonate With another user? | Example |
|-------|-----------------------------------|---------------------------------------|---------|
| Level 1 | Yes | Yes | IBE |
| Level 2 | No | Yes | CL-PKC |
| Level 3 | No | No | CB-PKC |
| Level 4 | No | No | SCS |

By the logic behind the PKC, the Private Key is secret to the user and the Public Key, publicly distributed over the network, need not be protected for confidentiality. It will make them subjected to several active attacks (such as the substitution of a "false" Public Key for a "true" one in a PKI directory). That is why, with the key pair and user's identity (ID), it is required to have the user's attribute (G), which guarantees that $usk_{Pr}$ is the original Public Key of the user.

## 3.4    Related work

This section includes the literature on approaches to solving the key escrow problem in Public Key encryption.

### 3.4.1   Threshold Key Issuing

Threshold Key Issuing (TKI) allows multiple parties to jointly generate and distribute a secret key without one party having full knowledge [7]. The idea behind TKI is to split the secret key into multiple shares, such that a certain number of shares are required to reconstruct the key. This number is known as the threshold and is typically chosen such that it is less than the total number of shares but greater than or equal to half of the shares. The shares are distributed to the participating parties, and any subset of parties with the threshold number of shares can reconstruct the secret key [62]. TKI can be implemented using various mathematical techniques, such as Shamir's Secret Sharing or Polynomial Secret Sharing. These techniques involve generating a polynomial over a finite field, with the secret key as the constant coefficient [54]. The polynomial is then evaluated at different points to obtain the shares distributed to the parties. TKI has several advantages over traditional key distribution methods, such as eliminating the need for a trusted third party to generate and distribute the key and improving the security of the key by distributing it among multiple parties. TKI is commonly used in applications that require high levels of security, such as in financial transactions, military communication, and secure multi-party computation.

### 3.4.2 Certificate-Based Cryptosystem

A Certificate-Based Cryptosystem (CBC) is a public key cryptography scheme in which the identity of a user is verified by a trusted third-party certificate authority (CA) using digital certificates [44, 45]. In a CBC, each user is assigned a public key by the CA, which is included in a digital certificate digitally signed by the CA. The digital certificate contains the user's identity, public key, and CA's digital signature. To send a message to another user, the sender encrypts the message using the recipient's public key, obtained from the recipient's digital certificate. The recipient can then decrypt the message using their private key, which is kept secret.

### 3.4.3 Certificate-Less Public Key Cryptosystem

Certificate-less Cryptosystems (CLC) are public key cryptography that addresses the key escrow problem by eliminating the need for digital certificates and trusted third-party Certificate Authorities (CAs) [46]. In CLC, users generate their own public and private keys without the involvement of a CA. The public key is derived from the user's identity information, such as an email address or a username, and a secure hash function. The user generates the private key using a random number generator or another secure process. The user's identity information and their public key are then distributed to other users who wish to communicate with them. By eliminating the need for digital certificates and trusted third-party CAs, CLC removes the possibility that a government or other entity may force a trusted third party to divulge a user's private key. Additionally, since the user generates the private key, there is no risk of a third-party holding a copy of the key.

However, there are some limitations to CLC. One of the main limitations is the difficulty of key distribution since the public key must be distributed through other secure channels, such as email or in-person exchange. Additionally, since there is no centralized authority to verify identities, there is a risk of spoofing or impersonation. Despite these limitations, CLC is becoming an increasingly popular option for secure communication and data sharing in situations where traditional certificate-based systems may not be feasible or desirable, such as in peer-to-peer networks or in situations where there is a lack of trust in centralized authorities.

### 3.4.4 Hierarchical Identity-Based Cryptosystem

Horwitz and Lynn [33] suggested a hierarchy of PKGs in identity-based encryption. First, they introduce the 2 Level hierarchical IBE, where total collusion resistance is obtained at the first level and partial collusion resistance at the second Level. So, it has only limited resistance to user collusion. In scheme [39], Gentry and Silverberg extend the Boneh and Franklin scheme [7] to construct the fully flexible Hierarchical ID-based encryption scheme. PKGs generate Private Keys only for the users immediately below them in the hierarchy, as shown in figure 3.4.

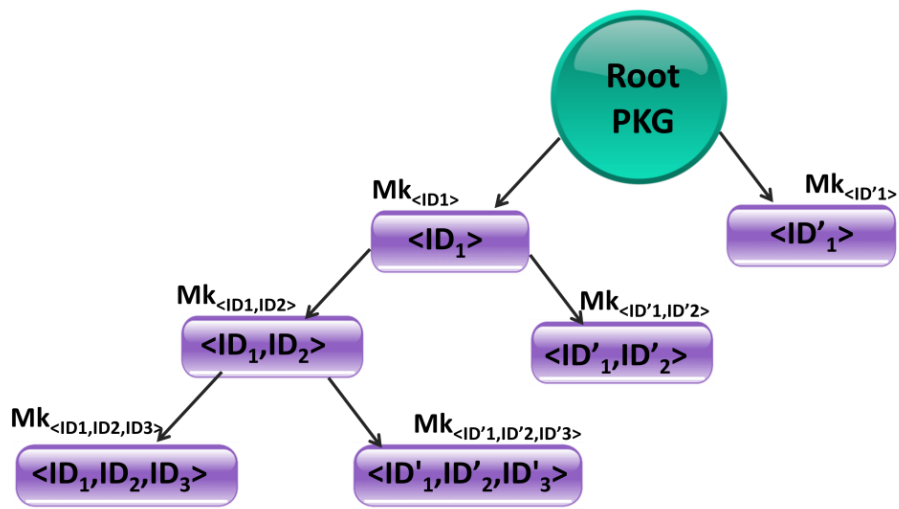

Figure 3.4  Hierarchy of Private Key Generators

A hierarchical identity-based cryptosystem (HIBC) enables the creation of a hierarchical structure for cryptographic keys. This structure allows for the delegation of authority within an organization or community, where individuals with higher levels of authority can access information protected by lower levels of authority. In a hierarchical identity-based cryptosystem, each user is identified by a hierarchical name that reflects their position in the organization. For example, a user might be identified as "John.Smith@Department1.Company.com". The user's hierarchical name derives the public key associated with this identity.

## 3.5 Comparison of existing ID-based encryption scheme

Boneh and Franklin proposed a solution to key escrow in which a master key is derived from the number of different Private Key Generators (PKG) so that each PKG only knows a partial secret of the Private Key, and no single PKG has complete knowledge of it [7]. However, this approach requires a significant amount of infrastructure and computational cost. Similarly, hierarchical identity-based and certificate-based encryption requires extra infrastructure and computational costs [33, 39]. To address the problem of key revocation, Al-Riyami et al. [46] successfully eliminated the certificate requirement in their scheme. However, their scheme only provides implicit authentication of the Public Key, which means that the sender cannot know for certain whether the Public Key is the real Public Key of the receiver. In VIBE [31], another Public Key and Private Key pair is used to solve the key escrow problem. Still, this scheme is slightly slower than certificate-less Public Key encryption due to the extra point addition operation it requires. Tables 3.2 and 3.3 compare Public Key encryption schemes that avoid key escrow problems based on computation cost and trust level.

Table 3.2 Comparison of Computation Cost of a variant of Public Key encryption which avoids key escrow problem, where M: Point multiplication, P: No of pairing operation, E: Exponentiation, t: no of user's identities in HIBE from root to leaf and n: number of PKGs in threshold issue scheme

| Scheme | Key Generation | Encryption | Decryption |
|---|---|---|---|
| Threshold key issue | 2nM | 1M+1P+1E | 1M+1P |
| CBC | 4M | 2M+1P+1E | 1P |
| CLC | 4M | 1M+1P+1E | 1M+1P |
| HIBE | 2tM | tM+1P+1E | tM +tP |

Table 3. 3 Comparison of public information and trust level to PKG of different schemes, where, $N_1 = usk_{Pr}P$: point multiplication of user secret $usk_{Pr}$ with group generator P, $N_2 = usk_{Pr}.msk_{Pub}$: point multiplication of user secret $usk_{Pr}$ with PKG's Public Key and C: commitment of user's secret key with the PKG's Public Key.

| Scheme | Public information | Trust level |
|---|---|---|

| Threshold key issue | ID | Level 1 |
|---|---|---|
| CB-PKC | ID, $usk_{Pub}$ | Level 3 |
| CL-PKC | ID, $usk_{Pub}$, C | Level 2 |
| HIBE | $ID_i$ | Level 1 |

## 3.6   Summary

This chapter summarises the fundamental building blocks of cryptography after presenting the basic mathematics in chapter 2. The first section of the chapter introduces public key encryption and its advantages and disadvantages, particularly in managing user certificates. The second section focuses on the thesis's main topic, identity-based encryption. It defines the basic concept of identity-based encryption and compares it with Public Key Infrastructure (PKI). It also discusses the applications, advantages, disadvantages, and security definition of identity-based encryption. The key escrow problem is introduced in the following section, along with a detailed discussion of some related literature on the subject. Finally, the chapter compares the existing schemes for solving the key escrow problem.

# Secure Key Issue

This chapter expertly designs an ID-based encryption scheme that serves as a fair balance between the users and government agencies. Here, "Fair balance" means the user has the right to privacy, and the government can monitor the user's criminal activity over secure communication. In the first section (section 4.1), we define some existing solutions regarding the key escrow problem that will be useful for describing the given proposed model. Section 4.2 describes the proposed scheme by defining cryptographic goals based on the earlier threat model. In the following section, we design decisions on achieving these objectives and how this impacts our model. Section 4.2 concludes with a concrete proposal in the form of an algorithm and an evaluation section motivating why our cryptographic design goals were successfully met. In the next section (section 4.3), we define the security model for our scheme, which includes IND-CCA Type I and Type II attacks. The mathematical implementation of our model is applied in section 4.4.

## 4.1   Models of existing solution

We already discuss several approaches [22, 31, 39, 44, 45, 46, 47, 48] that avoid the Key escrow problem inherited in the generic IBE scheme [5] in section 3.4. The most common work on Identity-based encryption to avoid key escrow problems is introduced by the Boneh-Franklin [7] using the threshold cryptography technique [50]. But it requires a large amount of infrastructure and more computational cost to implement such a scheme. Likewise, the same drawbacks exist with the hierarchal IBE [9, 12], where the tuple of identities identifies the public key user from roots to him. Certification-based encryption [45] does not preserve the identity-based encryption property. The public key revocation is additionally a significant disadvantage. However, the public key revocation is not a big issue; it can be removed by managing the key revocation list. That becomes a new problem and may require a high volume of space to store the certificates and computational time to verify those certificates. To solve the problem of managing revocation list, certificate-less encryption [46] was introduced. Indeed, this public key encryption variant solves many problems: key escrow problem, public key revocation, management of certificates, etc. it provides only implicit authentication to the public key. That means the sender will never know whether the given public key is the original public key of the recipient.

## 4.2    Proposed Model: M-IBE

This section describes the cryptographic requirements for designing the proposed model.

### 4.2.1   Security Goals

The standard general goals stated that the proposed scheme should be user-friendly and efficient. In addition to these generals' goals, it is also to define some cryptographic goals.    A well-made encryption scheme should be able to achieve the following cryptographic goals:

- **Authenticity**: The recipient has reasonable assurance that the claimed sender sends the message.
- **Confidentiality**: The message is protected from disclosure by unauthorized persons.
- **Integrity**: The message is protected from being modified by unauthorized users.
- **No key escrow**: The user's private keys are only disclosed to the owner of the claimed identity. No other user should be able to retrieve the private key.
- **Key validation**: Each user in the system should be able to verify the correctness of their private key.
- **Limited key validity**: The user's private key should not be valid for a limited period.

### 4.2.2   Design to Achieve Goals

In this section, our model is modified according to achieving the following cryptographically goals:

*Authenticity:*

Authenticity can be achieved by depending on authenticated encryption scheme. The authentication mechanism still depends on the security guarantees of the IBE scheme. Since a trusted third party, known as a private key generator, verifies the user's unique identifier corresponding to their public key. Accordingly, such an IBE confirms that a message encrypted with a public identifier can only be seen by the corresponding private key. If the authentication mechanism is insufficient, thus anyone could use it to impersonate the user. Our proposal is based on the Boneh-Franklin IBE scheme. Therefore, the scheme achieves authenticity.

*Confidentiality:*

Confidentiality can be achieved by applying an encryption scheme before sending a message. Identity-based encryption (IBE) can reach confidentiality and the general design goals of usability and applicability. During the design of our scheme, we can consider having several IBE schemes: Boneh and Franklin IBE [7], Sakai and Kasahara IBE [59] and Gentry IBE [37]. For the convenience of the desirable issue, we use the Boneh-Franklin IBE scheme as the encryption scheme.

*Integrity:*

Similar to authenticity, integrity can also be achieved depending on the security guarantees of the IBE scheme. If the scheme is sufficiently authenticated, no one can impersonate the user. As discussed in our proposal, our proposed scheme is derived from the Boneh-Franklin scheme.

*No Key Escrow:*

The IBE scheme inherently implies a property known as key escrow, which is undesired in the most practical system. To bypass the key escrow problem, multiple PKG was implemented as a distributed key generation system for IBE and several other schemes discussed in Section 3.4. To avoid the key escrow problem, we have proposed a scheme that will be discussed in section 4.2.4.

*Key validation:*

Key validation can be achieved by using the secret information to lock the partial private key on a trusted third party and unlock it using his secret information. Each can verify the private key with the use of some parameter, as shown in Algorithm 4.1

*Limited key validity*

IBE scheme does not provide the revocation of the public key facility in the generic scheme. To attain the revocation of the public key, we can embed an expiration date along with the user identity ID. Thus, as a part of the public key, the expiration date should be publically available to everyone.

### 4.2.3   Proposed Model

To bypass the key escrow problem inherited in the IBE scheme, an additional entity known as a private key privacy organization (PKPO) is added to our model. For convenience, we call our model M-IBE. Before explaining the model, some definition should be required to understand our model.

**Definition 4.1: (Key escrow)** In Identity-based encryption, the user's private key is generated and stored in PKG. This unusual property inherited in IBE is called the key escrow.

**Definition 4.2: (Key escrow Problem)** An unusual property inherited in a generic Identity-based encryption scheme allow:

1. PKG to use the user's private key in a mischievous activity without their permission and pretend as a user.
2. The user may deny that message is not sent by him and claim to PKG.

This situation is called the key escrow problem.

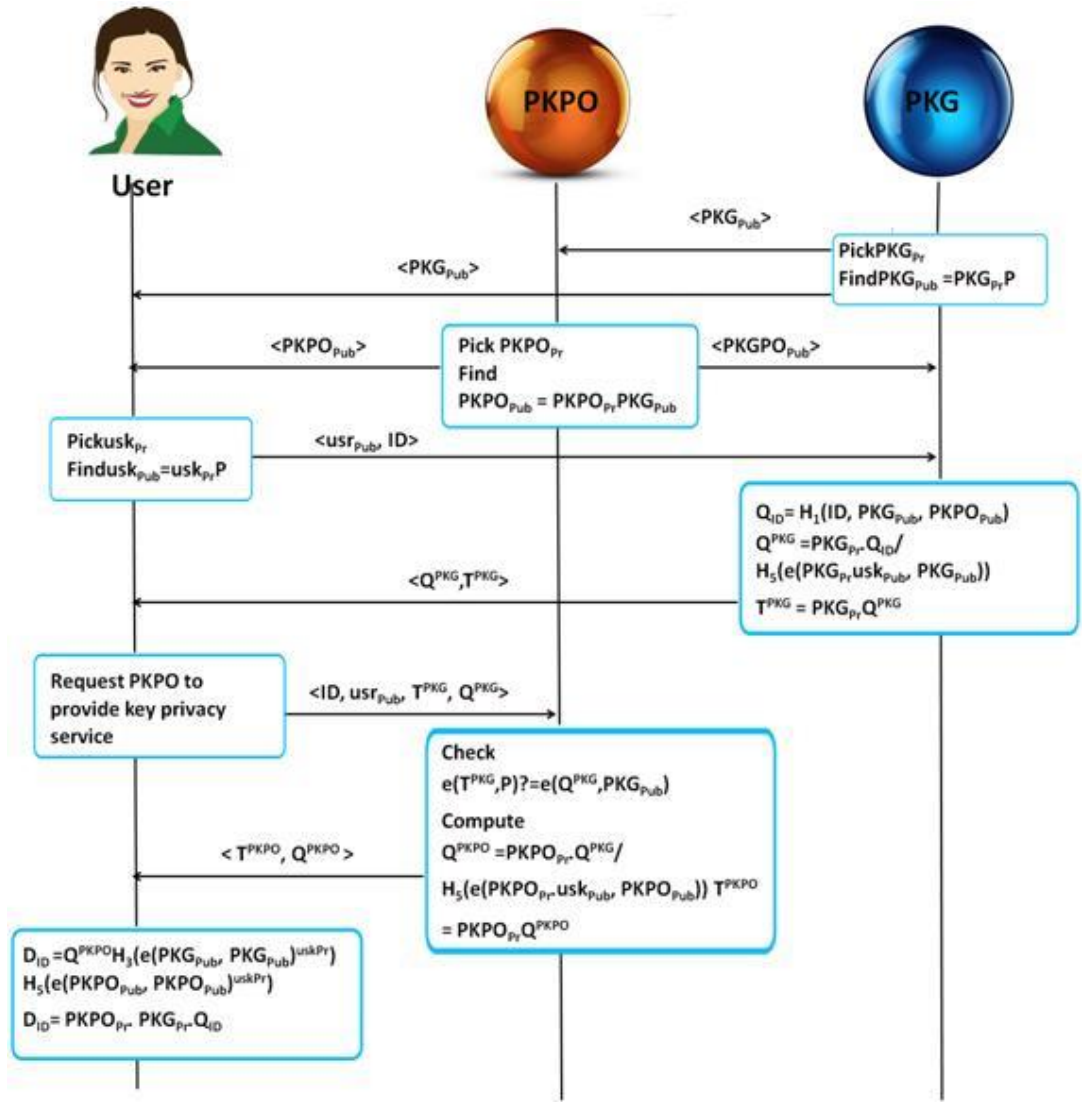

Figure 4. 1 Steps for Private Key Issue

**Definition 4.3: (Private Key Generator)** A single PKG to check the user's identification and provide a partial private key to the user.

**Definition 4.4: (Private Key Privacy Organization)** A single key privacy agency is a Non-Government Organization who has works between the user and PKG (Government agency) for the sake of users. PKPO is introduced to provide the privacy service to the private key by providing their signature in a confused manner.

**Definition 4.5 (Judge)** In case of malicious PKG, PKPO or user, anyone can file a case in court, such that the judge recovered the private key in the presence of all three entities. Consequently, monitor the malicious communication.

**Definition 4.6: (User)** An entity who wants privacy in communicating with another entity over an insecure medium.

**Assumption:** Here, we assume that PKG and PKPO are never colluding so that malicious PKG and PKPO can never use their private key.

This Model consists of two algorithms (Setup and Key Extract). Further, the KenGen algorithm is divided into three processes (Partial key issuing, Key securing process and Key Fetching). First, PKG runs the Setup algorithm to create the $<pkg_{Pr}, pkg_{Pub}>$, sequentially PKPO generates his key pair $<pkpo_{Pr}, pkpo_{Pub}>$ and publish public parameters keeping their secret key to themself. To get the partial key, the user requests PKG by sending his identity and hash of his secret info as a confusing factor. PKG checks the user identity and provides the partial private to the user in a confused manner if and only if the user is legitimate. Now, the user requests PKPO to provide privacy service, and PKPO returns the original private key. Only the legitimate user with a secret key for a confusing factor unlocks the message to get the original private key. The user fetches the original private key if he has a secret key corresponding to unlocking the confusing factor. For a given user identity and message, the user encrypts/signs the message. The user can decrypt the message for a particular Ciphertext and his private key.

### 4.2.4 Secure Key Scheme

This section proposes a model for Identity-based encryption free from key escrow problems, as shown in Algorithm 4.1. This model consists of four algorithms (Setup, Key Extract, Encrypt, and Decrypt).

---

**Algorithm 4.1 Acquire secure private key**

---

**Aim**: Alice wants to request the PKG to acquire his private key securely.

**Result**: PKG and PKPO are issuing a private key.

1. **Setup**: Let $P$ the generate the Group $G^*$. Given $e: G_1 \times G_1 \to G_T$ is bilinear

mapping, $H_1: \{0,1\}^* \to G^*$, $H_2: F^* \to \{0,1\}^l$, $H_3: \{0,1\}^n \times \{0,1\}^n \to Z_q$, $H_4: \{0,1\}^n \to \{0,1\}^n$, and $H_5: G_2 \to Z_q$ are five hash functions, where $l$ denotes the length of a message. The PKG randomly chooses a master key $pkg_{Pr} \in Z_q$ and generate his public key $pkg_{Pub} = pkg_{Pr}.P$. And then, KPA randomly chooses a key $pkpo_{Pr} \in Z_q$ and create his public key $pkpo_{Pub} = pkpo_{Pr}.pkg_{Pub}$. Now, publicly distribute the parameter $<G, F, H_1, H_2, H_3, H_4, H_5, pkg_{Pub}, pkpo_{Pub}>$.

2. **Key Extract:** As shown in figure 4.1, three entities (user, PKG, and PKPO) are participating in issuing a private key. This process includes the following three stages:

   ▪ **Partial Key Supply**: The user chooses $uskg_{Pr} \in Z_q$ and generate his public key $usk_{Pub} = usk_{Pr}P$ and request PKG to provide a partial private key by giving $uskg_{Pub}$ and ID as follows:

      ▪ Check the identification of users

      ▪ Compute the public key of a user as
      $$Q_{ID} = H_1(ID, P_0, P_1)$$

      ▪ Compute Partial private key
      $$Q^{pkg} = \frac{pkg_{pr}Q_{ID}}{H_5\big(e(pkg_{Pr}usk_{Pub}, pkg_{Pub})\big)}$$
      $$T^{pkg} = \text{pkg}_{\text{Pr}}Q^{\text{pkg}}$$

      ▪ Moreover, it sends it to the user

   ▪ **Key Securing**: The user requests PKPO to provide key privacy service by sending ID, $usk_{Pub}$, $T^{pkg}$ and $Q^{pkg}$

      ▪ Check $e(T^{pkg}, P) == e(Q^{\text{pkg}}, \text{pkg}_{\text{Pub}})$

      ▪ Compute $Q^{\text{pkpo}} = \frac{\text{pkpo}_{\text{Pr}}Q^{\text{pkg}}}{H_5\big(e(\text{pkpo}_{\text{Pr}}usk_{Pub}, \text{pkg}_{\text{Pub}})\big)}$
      $$T^{pkpo} = \text{pkpo}_{\text{Pr}}Q^{\text{pkpo}}.$$

      ▪ Send $T^{pkpo}$ and $Q^{\text{pkpo}}$ to the user.

   ▪ **Key Fetching**: The user retrieves his private key

$$D_{ID} = Q^{pkpo}H_5(e(\text{pkg}_{\text{Pub}}, \text{pkg}_{\text{Pub}})^{\text{usk}_{\text{Pr}}}).H_5(e(\text{pkpo}_{\text{Pub}}, \text{pkpo}_{\text{Pub}})^{\text{usk}_{\text{Pr}}})$$

$$D_{ID} = \text{pkg}_{\text{Pr}}.\text{pkpo}_{\text{Pr}}Q_{ID}$$

The user can check the correctness of his private key by $e(Q_{ID}, pkg_{Pub})$ ?$= e(D_{ID}, P)$.

## 4.3 Security Model

Now, we are ready to define the adversaries for the M-IBE scheme. The general security definition for IBE requires the indistinguishability of encryptions against a fully-adaptive chosen ciphertext attacker (IND-CCA). By this definition, we have two entities, the adversary $Adv$ and the challenger X. After presenting the random public key, the adversary controls it in three phases. In phase 1, $Adv$ randomly constructs decryption queries on the Ciphertext. In the challenge phase, $Adv$ choose $M_0, M_1$ and C* for messages $M_b$ given by the challenger, where $M_0$, and $M_1$ are two random messages and C* is challenged Ciphertext. Phase 2 continues constructing more decryption queries; indeed, it cannot have info for the decryption of C*. Finally, $Adv$ guess bit $b'$ corresponds to $b$. The $Adv$ 's advantage is defined to be

$$Adv(Adv) = 2\left(Pr[b' = b] - \frac{1}{2}\right)$$

Here, we explore the BF model to permit adversaries to extract the partial private keys, or private keys, or both, for random Identities and replace the public key with identity with a random value.

A list of activities that an adversary can take against an M-IBE scheme is given below:

- *Partial Key Supply*: To derive the partial key $< Q^0, T^0 >$ for user $Adv$, X provides the output by running the algorithm Partial-key-supply.

- *Key securing*: To acquire the mystified private key $Q^1$ for user $Adv$, X gives output by running the algorithm key-securing.

- *Key obtaining*: $Adv$ requests the user's private key. To compute the actual private key X , the algorithm key extracts if the corresponding public key is not changed.

- *Request public key*: Suppose $Adv$ has public keys. To calculate the public key $P_A$ for user $Adv$, $C$ runs the algorithm set-public-key and responds to $Adv$.

- *Replace public key*: $Adv$ can adaptively replace the public key $P_A$ for user $Adv$ with any random $P'_A$.

- *Decryption query*: To get private key $S_A$, X runs the algorithm Set-Private-Key and then the decryption algorithm and respond to $Adv$, if $Adv$ has not substituted the user's public key. Otherwise, X could not decrypt. However, our need is that X decrypts Ciphertexts for those public keys have been substituted. However, $Adv$ is permissible to substitute the public key for $ID_{ch}$ with a new ID and then request a decryption of C*.

We assumed that adversaries who have master-key were not permitted to substitute public keys. Here, we will try our PKG to achieve the same level of trust as CA in a traditional PKI. So we will classify adversaries into two types with different potentials:

**M-IBE Type I Adversary:** Denoted as $A_I$, such adversaries do not have master-key. Indeed, $A_I$ can request public keys and substitute them with new values of its choice, extract partial private and private keys and constructs decryption queries for each identity of its choice. Additionally, some limitations on adversary $Adv_I$ are:

1. Given $ID_{ch}$, $Adv_I$ cannot extract the private key.

2. If the user's public key has already been substituted, $Adv_I$ cannot request the private key for any identity.

3. Before the challenge phase, $Adv_I$ do not allow substituting the public key for the challenge identity $ID_{ch}$ and extracting the partial private key.

4. In Phase 2, $Adv_I$ does not allow the construction of a decryption query on the challenge Ciphertext $C_{ch}$ with an identity $ID_{ch}$ and public key $P_{ch}$.

**M-IBE Type II Adversary:** Denoted as $Adv_{II}$, such adversaries have master-key. Indeed, $Adv_I$ cannot substitute the public key. Using the master key, Adversary $Adv_{II}$ can compute partial private keys. Additionally, some limitations on adversary $Adv_I$ are:

1. $Adv_{II}$ does not allow substituting public keys.

2. $Adv_{II}$ does not allow extracting the private key for $ID_{ch}$.

3. In Phase 2, $Adv_I$ do not allow the construction of a decryption query on the challenge Ciphertext $C_{ch}$ with an identity $ID_{ch}$ and the public key $P_{ch}$.

**Chosen Ciphertext security for M-IBE**

M-IBE scheme is semantically secure against an adaptive chosen Ciphertext attack ('IND-CCA secure") if no polynomial bounded adversary A of Type I or Type II has a non-negligible advantage against the challenger in the following game:

- Setup: For security parameter K, Challenger X runs the Setup algorithm. It responds $Adv$ to the output of system parameters $params$. Challenger X keeps the master key for Type I adversary. Otherwise, it gives to $Adv$.

- Phase 1: $Adv$ provides the number of requests. Each request for partial private key extraction, a private key extraction, a public key, a substitute public key or a decryption query for an individual user. According to the rules defined above, these queries can be run adaptively.

- Challenge Phase: $Adv$ responds to the challenge identity $ID_{ch}$ and two equal-length message $M_0$ and $M_1$. The Challenger randomly chose a bit b $\in \{0, 1\}$ and computes $C_{ch}$. If encryption gives $\perp$, then $Adv$ lose the game. Otherwise, $C_{ch}$ is given to $Adv$.

- Phase 2: According to the rule defined above, A provides a second sequence of requests similar to Phase 1.

- Guess: A response with a guess $b' \in \{0, 1\}$. If $b' = b$, $Adv$ wins the game with an advantage $Adv(A) = 2\left(Pr[b' = b] - \frac{1}{2}\right)$

## 4.4    M-IBE schemes from Pairing

In this section, we implement identity-based encryption free from key escrow problems based on our proposed model. The first scheme, BasicM-IBE, is identical to the BasicIdent scheme of [7] and contains the basis of our most important scheme FullM-IBE. The master scheme is identical to the scheme FullIdent of [7]. Assuming the difficulty of GBDHP, our scheme is IND-CCA secure.

### 4.4.3    Basic M-IBE scheme

Here, we describe the four algorithms required to understand the fundamental ideas underlying our scheme that is identical to BasicIdent [7].

<div style="border:1px solid black; padding:10px">

**Algorithm 4.2 Identity-based encryption free from key escrow problem**

**Aim**: Given Bob's identity, $ID_B$, and private $D_{ID}$ obtain from algorithm 4.1, Alice wants to send an encrypted message to Bob so that no other (including PKG) than Bob can decrypt the message.

**Output**: Alice sends an encrypted Ciphertext $C$ successfully decrypted by Bob without losing confidentiality.

1. **Setup**: This phase is identical to the Setup phase of Algorithm 4.1.

2. **Key extract**: Identical to the Key extract phase in algorithm 4.1.

3. **Encrypt**: Alice randomly chooses $r \in Z_q$ and may encrypt her message using Bob's identity ID by

   - Compute $Q_{ID} = H_1(ID, pkg_{Pub}, pkpo_{Pub})$ and $g = H_2\big(e(Q_{ID}, pkpo_{Pub})\big)$

   - The resulting Ciphertext $C = <U, V> = <rP, M \oplus H_2(g^r)>$

   - C sent it to Bob.

4. **Decrypt**: Bob can decrypt $C$ by computing

   $$g' = e(U, D_{ID}) \text{ and } M = V \oplus H_2(g')$$

</div>

The scheme's consistency will be discussed in the next chapter, and we will analyze the value $g^r$ in encryption is similar to the $e(U, D_{ID})$ in decryption. This completes the BasicM-IBE scheme.

### 4.4.4 Full M-IBE scheme

To convert the BasicM-IBE scheme to the FullM-IBE scheme, this is a chosen Ciphertext secure IBE system in the ROM [58]. Taking all the cryptographic goals, IBE based on our model is presented in Algorithm 4.3.

<div style="border:1px solid black; padding:10px">

**Algorithm 4.3 Identity-based encryption free from key escrow problem**

**Aim**: Given Bob's identity, $ID_B$ and private $D_{ID}$ obtain from algorithm 4.1, Alice wants to send an encrypted message to Bob so that no other (including PKG) than

</div>

Bob can decrypt the message.

**Output**: Alice sends an encrypted Ciphertext $C$ that Bob successfully decrypts without losing confidentiality.

1. **Setup**: This phase is identical to the Setup phase of Algorithm 4.1.

2. **Key extract**: Identical to the KeyExtract phase in algorithm 4.1.

3. **Encrypt:** Alice randomly chooses $r \in Z_q$ and may encrypt her message using Bob's identity ID by

   - Compute $r = H_3(z, M)$, where $z \in \{0,1\}^*$
   - $Q_{ID} = H_1(ID, pkg_{Pub}, pkpo_{Pub})$ and $g = H_2\big(e(Q_{ID}, pkpo_{Pub})\big)$
   - The resulting Ciphertext $C = <U, V, W> = <rP, z \oplus H_2(g^r), M \oplus H_4(z)>$
   - C Sent to Bob.

4. **Decrypt**: Bob can decrypt $C$ by computing

   - $g' = e(U, D_{ID})$ and $z' = V \oplus H_2(g')$
   - $M' = W \oplus H_4(z')$ and $r = H_4(z', M)$

   Return the message $M'$ if U= $r'P$.

The scheme's consistency will be discussed in the next chapter, and we analyze that in decryption, z' and encryption z are equals. This completes the FullM-IBE scheme.

## 4.5 Summary

This chapter presents the implementation of the scheme proposed in section 1.1, which aims to address the key escrow problem in cryptographic systems. Firstly, a review of the related scheme model is provided. Subsequently, cryptographic objectives that must be achieved in the proposed scheme are defined, and a model is designed to meet these objectives. In the subsequent section, an algorithm for the proposed scheme is described, and BasicM-IBE and FullM-IBE are implemented, which are similar to BasicIden and FullIden as presented in [7]. The Full-IBE is designed to be secure against IND-CCA adversaries.

# Consistency, Security Proof, and Performance Analysis

In this chapter, first, we verify the consistency of Algorithm 4.2 and Algorithm 4.3. Then, we prove that our FullM-IBE scheme is secure against IND-CCA1 and IND-CCA2 adversary attacks. In the following section, we compared our scheme with the existing scheme. Finally, we claim that our scheme fulfils the property discussed in section 5.4.

## 5.1 Consistency of the M-IBE scheme

Here, we prove the correctness of the encryption and decryption stage of Algorithm 4.1 and Algorithm 4.2

### 5.1.1 BasicM-IBE scheme

The consistency of the scheme is verified as follows:

We know that $g = e(Q_{ID}, pkpo_{Pub})$

So
$$g^r = e(Q_{ID}, pkpo_{Pub})^r$$

$$= e\left(Q_{ID}, pkpo_{pr}pkg_{pr}P\right)^r$$

$$= e\left(pkpo_{pr}pkg_{pr}Q_{ID}, P\right)^r$$

$$= e(pkpo_{pr}pkg_{pr}Q_{ID}, rP)$$

$$= e(D_{ID}, U)$$

Here, we notice that the value $g^r$ in encryption is similar to the $e(D_{ID}, U)$ in decryption.

### 5.1.2 FullM-IBE scheme

The consistency of the protocol easily verified from

$$g' = e(D_{ID}, U)$$

$$= e(pkpo_{pr}pkg_{pr}Q_{ID}, rP)$$

$$= e\left(pkpo_{pr}pkg_{pr}Q_{ID}, P\right)^r$$

$$= e\left(Q_{ID}, pkpo_{pr}pkg_{pr}P\right)^r$$

$$= e\left(Q_{ID}, pkpo_{pub}\right)^r$$

$$= g^r$$

Therefore, in decryption, $g'$ and encryption $g^r$ is equal. Consequently, applying decryption on a ciphertext recovers the original message m. Moreover, our scheme achieves some influential properties that differentiate it from existing ID-based cryptosystems.

## 5.2 Secure against IND-CCA.

This section proves that our scheme is secure against the IND-CCA type I and II attacks. Public Key encryption scheme HybridPub [46] will be used as a tool in security proof of FullM-IBE.

**Theorem 1**: Consider four Random Oracle hash functions $H_1, H_2, H_3, H_4$ and $H_5$. M-IBE is IND-CCA secure if there is no bounded polynomial algorithm [19] that can solve the GBDHP in groups generated by $G$ with non-negligible advantage.

**Proof**: This theorem is similar to Theorem 1 in [46]. Theorem 1 can be proved by proving the number of lemmas. It can be made into a concrete security reduction relating the advantage ϵof Type I or Type II attackers against M-IBE to that of an algorithm to solve GBDHP or BDHP. Theorem 1 for Type I adversaries follows by combining Lemmas 2, 3 and 4. Similarly, Theorem 1 for Type II adversaries combines Lemmas 4, 5 and 6.

**Lemma 2**: Consider there are five Random Oracle hash functions $H_1, H_2, H_3, H_4$ and $H_5$, and there has been an adversary $Adv_I$ of IND-CCA Type I against FullM-IBE with advantage $\epsilon$, running time $t$. Suppose A constructs at most $q_i > 0$ queries for $H_i$, where $1 \leq i \leq 5$ and at most $q_D > 0$ queries to decryption. Then there is an adversary B that behaves either as a Type I or a Type II IND-CPA adversary and has the advantage of at least $\frac{\epsilon}{4q_1q_5}\lambda^{q_D}$ against HybridPub. Its running time is $t + O(q_3 + q_4)q_d t'$. Where,

$$1 - \lambda \leq (q_3 + q_4).\epsilon_{OWE}(t + O(q_3 + q_4)q_d t', q_2)$$
$$+ \epsilon_{GBDHP}(t + O(q_3 + q_4)q_d t' + 3q^{-1} + 2^{-n+1})$$

The advantage of any type I or Type II is at least $\epsilon_{OWE}(T, q')$, runs in time $T$ and constructs $q'$ queries to $H_2$, and the advantage of any algorithm to solve GBDHP is at least $\epsilon_{GBDHP}(T)$. Here, $t'$ is the running time of the BasicM-IBE encryption algorithm.

**Lemma 3**: Consider $H_3$ and $H_4$ are Random Oracles. There is a Type I and Type II IND-CPA adversary $Adv_I$ and $Adv_{II}$, respectively, against HybridPub with advantage $\epsilon$ and construct at most $q_3 > 0$ queries to $H_3$ and at most $q_4 > 0$ queries to $H_4$. Then there is a Type I and Type II OWE adversaries $Adv_I$ and $Adv_{II}$, both have an advantage of at least $\frac{\epsilon}{2(q_3 + q_4)}$ against BasicPub. Its running time is $O\big(time(Adv_I)\big)$ and $O\big(time(Adv_{II})\big)$, respectively.

**Lemma 4**: Consider $H_2$ is a Random Oracle, and there is $Adv_I$ and $Adv_{II}$ of Type I and Type II OWE adversary that has an advantage $\epsilon$ against BasicPub, constructs at most $q_2 > 0$ queries to $H_2$. Then there is an adversary B to solve the GBDHP has the advantage at least $(\epsilon - \frac{1}{2^n})/q_2$. Its running time is $O\big(time(Adv_I)\big)$ and $O\big(time(Adv_{II})\big)$, respectively.

**Lemma 5**: Consider $H_1$ is a Random Oracle and that there is an IND-CCA Type II adversary $Adv_{II}$ on FullM-IBE with an advantage $\epsilon$ which makes atmost $q_1 > 0$ queries to $H_1$. Then there is an IND-CCA Type II adversary on HybridPub with an advantage of at least $\frac{\epsilon}{q_1}$ which runs in time $O\big(time(Adv_{II})\big)$.

**Lemma 6**: Consider $H_3$ and $H_4$ are Random Oracles, and there is a Type II IND-CCA adversary $Adv_{II}$ against HybridPub with advantage $\epsilon$ and construct at most $q_D > 0$ queries to decryption, at most $q_3 > 0$ queries to $H_3$ and most $q_4 > 0$ queries to $H_4$. Then there is a Type II OWE adversary $Adv_{II}$ against BasicPub with

$$time(Adv'_{II}) = time(Adv_{II}) + O(n(q_3 + q_4))$$

$$adv(Adv'_{II}) \geq \frac{1}{2(q_3 + q_4)}.((\epsilon + 1)(1 - q^{-1} - 2^{-n})^{q_D} - 1)$$

**Lemma 7**: In Lemma 2, Algorithm $\mathcal{KE}$ responds with correct to each decryption query with the advantage of at least $\lambda$ where

$$1 - \lambda \leq (q_3 + q_4).\epsilon_{OWE}(t + O(q_3 + q_4)q_d t', q_2)$$
$$+ \epsilon_{GBDHP}(t + O(q_3 + q_4)q_d t' + 3q^{-1} + 2^{-n+1})$$

Here, the advantage of any Type I or Type II is at least $\epsilon_{OWE}(T, q')$, operates in time $T$ and constructs $q'$ queries to $H_2$, and the advantage of any algorithm to solve GBDHP is at least $\epsilon_{GBDHP}(T)$. Here, $t'$ is the BasicM-IBE encryption algorithm's running time, and $t$ is adversary I's running time.

**Proof 2**: Let $Adv_I$ be a Type I IND-CCA adversary against FullM-IBE, has advantage $\epsilon$, runs in time t, construct at most makes $q_i > 0$ queries to Random Oracle $H_i (1 \leq i \leq 4)$, and a decryption query is at most $q_D > 0$. Another adversary $Adv_I$ derived from I pretends as a Type I IND-CCA adversary or Type II IND-CCA adversary. We assume two challengers' $X_I$ and $X_{II}$ are available B for two different challenges.

First, Adversary B chooses a Random bit c and an index I such that $1 \leq I \leq q_1$. B wishes to play with $X_I$ and aborts $X_{II}$ where $X_I$ passes B with a public key $K_{pub} = < G_1, G_2, e, n, P, P_0, P_1, Q, H_2, H_3, H_4 >$ if c = 0. Otherwise, B wishes to play with $X_{II}$ and aborts $X_I$ where $X_{II}$ passes B with a public key $K_{pub}$ along with the value $s_0$ and $s_1$ such that $P_0 = s_0 P$ and $P_1 = s_0 s_1 P$. Let the event that $Adv_I$ picks $ID_I$ such that $ID_I = ID_{ch}$ is denoted by $H$, the event that $Adv_I$ extract the partial private key for user $ID_I$ be $\Phi_0$ and the event that $A_I$ substitute the public key of user $ID_I$ denoted as $\Phi_1$.

Here, $H_1$ is a Random Oracle that will be ruled by B and managed as follows:

- $H_1$ queries: B manages a list for $H_1$ of tuples $< ID_i, Q, b_i, x_i, y_i, P_{0i}, P_{1i}, P_{2i} >$, empty in initially. When $Adv_I$ make queries $H_1$ on input $ID$, B outputs as follows:

a) B outputs $Q_1(ID) = Q_i$ if ID found in the $H_1$ list,

b) B Randomly chooses $b_I$ from $Z_q$, gives $H_1(ID) = b_I Q$ and adds the tuple $<ID_i, Q, b_I, \perp, \perp, x_i, y_i, P_0, P_1, P_2>$ to the $H_1$ list, if ID does not find in the $H_1$ list such that ID is I-th distinct $H_1$ query made by $Adv_I$.

c) Otherwise, when ID does not exist in the list and ID is the i-th distinct $H_1$ query made by $Adv_I$ where i $\neq$ I, B randomly chooses $b_i$, $x_i$ and $y_i$ from $Z_q$, gives $H(ID) = b_i P$ and adds $<ID, b_i P, b_i, x_i, y_i, x_i P, x_i P_0, y_i P>$ to the $H_1$ list if ID does not find in list where ID is the i-th $H_1$ query such that $I \neq i$.

After the definition of $H_1$, the FullM-IBE partial Private Key for $ID_i = b_i P_0$ such that $i \neq I$. Thus, the Public Key is $<x_i P, x_i P_0>$, and the Private Key is $<x_i b_i P_0>$ for $ID_i$ when c = 0. Otherwise, B can compute $s_0 b_I Q$, the partial Private Key of $ID_I$.

- $H_2$ and $H_4$ queries: $A_I$ makes $H_2$ queries and passed them to the challenger X. Adversary $A_I$ makes $H_4$ query and B passes this query to X, indeed, keeps list $<\sigma'_i, H_{4,i}>$ and X's answer to them.

- $H_3$ queries:

  1) Adversary $A_I$ makes $H_4$ query and B passes this query to X, indeed, keeps list $<\sigma'_i, M_j, H_{3,j}>$ and $C's$ answer to them.

  2) B manages a list of tuples $< Q, b_i, x_i, y_i, N_{1i}, N_{2i}>$ which is initially empty.

- $H_5$ queries:

  Let $Z_i = e(x_i y_i P, -x_i b_i P)$ When $A_I$ queries $H_5$, B responds as follows:

  a) If Z find in the $H_5$ list in tuple $<Z_i, Q_i, b_i, x_i, y_i, N_{1i}, N_{2i}>$, then B responds with $H_5(Z) = Q_i \in G_1^*$.

  b) If Z does not find in the list and Z is the I-th distinct $H_1$ query made by $A_I$, then B picks $b_I$ at Random from $Z_q$, outputs $H(Z) = b_I Q$ and adds the entry $<ID, b_I Q, b_I, \perp, \perp, N_1, N_2>$ to the $H_1$ list.

  c) Otherwise, B picks $b_I$, $x_I$ and $y_I$ at random from $Z_q$, output $N_1 = x_I b_I Q$ and $N_2 = x_I y_I P$ and adds the entry $<ID, b_I Q, b_I, \perp, \perp, N_1, N_2>$ to the $H_5$ list.

Phase 1: $A_I$ receives *params* from B and make many requests for a user, including a partial Private Key extraction, a private key extraction, a request for a public key, a substitution a Public Key or a decryption query. B replies to these requests as follows:

- Partial Private Key Extraction: Let $A_I$ request $ID_i$. One of the three cases will occur:

  - B outputs with $b_i P_0$, if $i \neq I$.

  - B aborts, if $i = I$ and $c = 0$.

  - B outputs with $s.b_I.Q$, if $i = I$ and $c = 1$.

- Private Key Extraction: Let $\mathcal{A}_I$ make the request on $ID_i$. Let the Public Key for $ID_i$ has not been the substitute. One of the two cases will occur:

  - $\mathcal{B}$ responds $x_i.b_i.P_0$, if $i \neq I$.

  - Otherwise, $\mathcal{B}$ aborts.

- Request for Public Key: Let $A_I$ request $ID_i$. B returns $<x_i P, x_i P_0>$ by obtaining the $H_1$ list.

- Replace Public Key: Let $A_I$ request $ID_i$ to substitute the Public Key with value $<P'_{0i}, P'_{1i}>$. One of the two cases will occur:

  - B aborts, if $i = I$ and $c = 1$.

  - Otherwise, the existing entries $P_{0i}$ and $P_{1i}$ in the $H_1$ list is substituted with the new entries $<P'_{0i}, P'_{1i}>$ and B requests X to substitute the Public Key $<P'_0, P'_1>$ in $K_{pub}$ with new values $<P'_{0i}, P'_{1i}>$, if $i = I$.

- Decryption Queries: Let $A_I$ request to decrypt the Ciphertext $< U, V, W >$ for $ID_l$, As B is pretending as an IND-CPA adversary, so he will not use the challenger X to reply to the query ( $I = l$ ). Alternatively, for existing Public Key $< P_{0I}, P_{1I} >$ of $ID_i$ and a ciphertext $C = <U, V, W>$, B responds to each decryption query with an advantage at least $\lambda$, where $\lambda$ is proved in Lemma 7. B responds to each decryption queries as follows:

  i. B search tuple $< \sigma_j, M_j\ H_{3,j} >$ on the $H_3$ list. Accumulate these tuples in a list $S_1$. If $S_1$ is empty, output $\perp$ and halt.

ii. B search every pairs < σ'$_i$, H$_{4,i}$> in the H$_4$ list, for every tuple < σ$_j$, M$_j$ H$_{3,j}$>in S1. If σ$_j$ = σ'$_I$, add tuple < σ$_j$, M$_{j,}$ H$_{3,j}$, H$_{4,i}$> inS$_2$ list. If S$_2$ is empty, then output ⊥ and halt.

iii. For W = M$_j$⊕H$_{4,i,}$, B find in S$_2$ for such an entry. If it exists, it responds Mj as the result of <U, V, W>. Otherwise, responds ⊥.

Challenge Phase: A response to the challenge identity ID$_{ch}$ and two equal length messagem$_0$, m$_1$∈ M. Let A$_I$ not allow extracting the Private Key for identity ID$_{ch}$. The algorithm Bresponds as follows:

- B aborts, if ID$_{ch}$≠ ID$_I$.

- Otherwise, ID$_{ch}$ = ID$_I$ and Bgives Xthe pair m$_0$, m$_1$ as the messages on which it wishes to be challenged.

X outputs C'= <U', V', W'>, such that C' is the HybridPub encryption of m$_b$. Then Bsets C*= $<b_I^{-1}$.U', V', W'> and passes C*to A$_I$ such that C*is the FullM-IBE encryption of m$_b$ with Public Key <P$_{0I}$, P$_{I1}$>. Let <P$_{0ch}$, P$_{1ch}$> be the Public Key for identity ID$_{ch}$ during the challenge phase.

Phase 2: Similar to phase 1, B repeatedly reply to A$_I$'s requests.

Guess: In the end, A$_I$ may form a guess b' for b. B responds b' as a guess for b. If A$_I$take more time than time t, or take many attempts to make q$_i$ queries or q$_D$ decryption queries, then B should abort A$_I$ and output a Random guess for bit b.

Analysis: During the whole execution process, if B does not abort and the decryption queries respond by B is uses correctly, then algorithm A$_I$ is considered to be a real attack. Furthermore, the encryption of m$_b$ is the challenge Ciphertext C* under the Public Key of ID$_{ch}$, such that b∈{0,1}is Random. So according to the definition of an adversary A$_I$ we have that 2(Pr[b = b']=1/2 ) ≥ ϵ.

Now we have to inspect that during the execution process the probability that B does not halt. Inspecting the execution process, we realize that B can abort for one of following reasons:

- When c = 0 and the event Φ$_0$ happens, denoted as H$_0$

- When $c = 1$ and the event $\Phi_1$ occurs, denoted as $H_1$
- Because $A_I$ made a Private Key extraction on $ID_I$ at some point represented as $\Phi_2$
- When $A_I$ chose $ID_{ch} \neq ID_I$, denoted as $\Phi_3$
- Or $A_I$ chose $Z_{ch} \neq Z_I$, denoted as $\Phi_4$

The probability that $ID_{ch} = ID_I$ is equal to $1/q_1$ because $A_I$ construct $q_1$ queries of $H_1$ and B have a choice to choose I from the set of $q_1$ queries. Hence $Pr[H] = Pr[\Phi_3] = 1/q_1$. Similarly, $A_I$ construct $q_5$ queries of $H_5$ and B can choose I from the set of the $q_5$ query. Thus, the probability that $Z_{ch} = Z_I$ is equal to $1/q_5$. Hence $Pr[\Phi_4] = 1/q_5$. As we know, if $A_I$ choose $ID_{ch} = ID_I$, then no Private Key extraction on $ID_I$ will be permitted. From all this information:

$$Pr[B \text{ does not abort}] = Pr[\neg H_0 \wedge \neg H_1 \wedge \neg \Phi_2 \wedge \neg \Phi_3 \wedge \neg \Phi_4] = \frac{1}{q_1}\frac{1}{q_5}Pr[H_0 \wedge H_1 \mid H]$$

Because two events $H_0$ and $H_1$ are mutually exclusive. So we can write

$$Pr[H_0 \wedge H_1 \mid H] = 1 - Pr[H_0 \mid H] - Pr[H_i \mid H]$$

And because the event $\Phi_i \mid H$ is independent of the event $c = I$

$$Pr[H_i \mid H] = Pr[(c = i) \wedge \Phi_i \mid H] = \frac{1}{2}Pr[\Phi_i \mid H]$$

Now, we have

$$Pr[B \text{ does not abort}] = \frac{1}{q_1 q_5}(1 - \frac{1}{2}Pr[\Phi_0 \mid H] - \frac{1}{2}Pr[\Phi_i \mid H])$$

According to the rules subjected to the adversary described in the security model, an adversary cannot allow extracting the partial private key or substitute the public key. So, we have $Pr[\Phi_0 \wedge \Phi_1 \mid H] = 0$. This implies that $Pr[\Phi_0 \mid H] + Pr[\Phi_1 \mid H] \leq 1$. Hence, we realize that $Pr[B \text{ does not abort}] \geq \frac{1}{2q_1 q_5}$

Finally, now we examine the probability that B in decryption query phase correctly controls all $q_D$ decryption queries of $Adv_I$. Thus, B's advantage is at least $\frac{\epsilon}{2q_1 q_5}\lambda^{q_D}$. It follows that either B's advantage as a Type I adversary or B's advantage as a Type II adversary $\frac{\epsilon}{4q_1 q_5}\lambda^{q_D}$. This completes the proof.

**Proof of Lemma 3**: This Proof is similar to the Proof of Lemma 10 of [53], modified by adding query $q_5$.

**Proof of Lemma 4**: This proof is identical to the proof of Theorem 4.1 in [7], with modification with the addition of query $q_5$ for type I adversary and Type II adversary.

**Proof of Lemma 6**: This lemma can be proven through theorem 14 of [53], assuming that msk$_{Pr}$ can be made available to Type II adversaries.

## 5.3    Performance

In chapter 4, we proposed the M-IBE scheme. Indeed, our approach is modifying the key generation process of the Boneh-Franklin scheme [7]. Table 5.1 shows the comparisons of our scheme with the existing scheme based on computation cost. According to the level of trust in PKG in Girault [22], Table 5.2 compares our scheme with different Public Key encryption schemes. Here, we conclude that our scheme achieves Level 3.

Table 5.1 Comparison of Computation Cost of the proposed scheme with the existing scheme, where M: Point multiplication, P: No of pairing operation, E: Exponentiation, t: no of user's identities in HIBE from root to leaf and n: number of PKGs in threshold issue scheme

| Scheme | Key Generation | Encryption | Decryption |
|---|---|---|---|
| Threshold key issue | 2nM | 1M+1P+1E | 1M+1P |
| CB-PKC | 4M | 2M+1P+1E | 1P |
| CL-PKC | 4M | 1M+1P+1E | 1M+1P |
| HIBE | 2tM | tM+1P+1E | tM +tP |
| VIBE | 4M | 1M+3P+1E | 2M+1P |
| M-IBE | 6M+1P+1E | 1M+1P+1E | 1M+1P |

Table 5.2 Comparison of public information and trust level to PKG of the proposed scheme with the existing scheme, where $N_1 = usk_{Pr}P$: point multiplication of user secret $usk_{Pr}$ with group generator P, $N_2 = usk_{Pr}.msk_{Pub}$: point multiplication of user secret $usk_{Pr}$ with PKG's Public Key and C: commitment of user's secret key with the PKG's Public Key.

| Scheme | Public information | Trust level |
|---|---|---|
| Threshold key issue | ID | Level 1 |
| CB-PKC | ID, $usk_{Pub}$ | Level 3 |

| CL-PKC | ID, $usk_{Pub}$, C | Level 2 |
| HIBE | $ID_1,...ID_i$ | Level 1 |
| VIBE | ID, $N_1$. $N_2$ | Level 3 |
| M-IBE | ID | Level 3 |

Besides our scheme, each scheme needs at most four hash functions, but our scheme needs one more extra hash function. However, the computational time of the hash function is not a big issue as it is swift compared to the pairing computation or scalar operation. So, we ignore the computational time of the hash operation and point addition because the number of hash operations is almost equal in all schemes and point addition is lightweight compared to other heavy operations. Based on tables 5.1 and 5.2, we compared our scheme with the existing scheme as follows:

### 5.3.1 Comparison with Boneh-Franklin

By including a Non-government organization (PKPO) between the user and PKG, we modify Boneh-Franklin's IBE scheme to remove the Key escrow Problem. PKPO provides key privacy to the user. The computation cost of our scheme's encryption and decryption algorithm equals the Boneh-Franklin scheme [7]. While in terms of Private Key issuing, our scheme needs two extra point multiplications, one pairing operation, and one exponentiation. Thus, for key issuing, our scheme is slightly slower. As compared to the solution of the key escrow problem solved by the Boneh-Franklin, our scheme is speedy in issuing the private key, as shown in Table 5.1. Compared with the BF approach, our scheme achieves level 3 of the trust level, as shown in Table 5.2.

### 5.3.2 Comparison with CL-PKC

Compared to CL-PKC, our scheme is slightly slower in the key issuing process but takes equal computation time in encryption and decryption. In our scheme, the user's secret information is used only to secure extracting the Private Key. Therefore, our scheme provides implicit and explicit authentication compared to CL-PKC, which only provides implicit authentication. Moreover, our scheme achieves level 3; conversely, CL-PKC achieves level 2 of trust level to PKG. From Table 5.2, we conclude that our scheme has at least one public information share on the communication network. At the same time, CL-PKC takes three public pieces of information.

### 5.3.3 Compared with other schemes

Compared with CB-PKC, our scheme preserves the advantage of ID-based cryptography and has one number of public information on the network. At the same time, CB-PKC has two numbers of public information, as shown in Table 5.2. Both schemes achieve the third level of trust on PKG. From a computational point of view, our scheme is faster in encryption and decryption. At the same time, our scheme is slightly slower than another scheme in the key issuing process.

Compared to VIBE, as shown in Tables 5.1 and 5.2, our scheme is performed well and faster in encryption algorithm and has a minimum number of public information shared on the network. Here, we analyze that our scheme is the most advantageous variant of the generic IBE scheme that removes the inherited key escrow problem with the following advantage:

- Minimum of public information shared on the network
- Achieve trust level 3 on PKG
- Preserve the advantage of the ID-based cryptosystem.

Moreover, with the minimum computational cost of encryption and decryption, our scheme will be more efficient for low-power consumption devices, where key issuing overload dominates the server side.

## 5.4 Claim

CLAIM 1: **No Key Escrow Problem.** *An adversary cannot decrypt the encrypted message without the knowledge of PKPO's Private Key* $pkpo_{pr}$; *even if they know the master key* $\mathrm{pkg_{pr}}$.

This claim follows from Theorem 1 in Section 5.2.

CLAIM 2: **Partial key escrow**. *An adversary cannot decrypt the encrypted message without the knowledge of PKPO's private key* $pkpo_{pr}$; *even if they know the master key* $\mathrm{pkg_{pr}}$.

This claim also follows from Theorem 1 in Section 5.2.

CLAIM 3: **Preserve ID-based cryptosystem**. *Compared to certificate-based encryption, where a certificate is required to obtain the Public Key, the user's identity ID is used to generate the Public Key.*

CLAIM 4: **Key recovery.** *In the case of PKG compromising the Private Key by the PKG, the user and the PKPO may claim that the Private Key $D_{ID}$ has been compromised by PKG and file a legal case against the PKG in court. Consequently, PKG will be present in the Court for their punishment. In the case of the malicious user, PKPO may claim on the victim's request that the user may be malicious and acting in a mischievous activity. Consequently, all three entities (PKG, PKPO, and user) will present in court on- a legal court order and release their master key to derive the user Private Key $D_{ID}$. Thus, the judge may recover the Private Key and decrypt the malicious message.*

## 5.5    Summary

This chapter demonstrates the consistency of the encryption and decryption algorithms 4.2 and 4.3. Furthermore, we prove that our scheme is secure against IND-CCA1 of type I and Type II attacks, as stated in Theorem 1, supported by Lemma 2 through Lemma 7. In the subsequent section, we perform a comparative analysis of our scheme with HIBE, VIBE, threshold key issuing, CL-PKC, and CB-PKC. Based on the results in Tables 5.1 and 5.2, our scheme outperforms the others. Finally, we assert that our scheme of transforming a generic IBE into a fair ID-based cryptosystem is consistent with the objective stated in Chapter 1 (Section 1.4).

# Conclusion and Future Work 6

This chapter provides a comprehensive review of the discussed topics and the outcome of the proposed scheme. Additionally, the limitations of our solution, possible future directions, and other domains in which our scheme is fruitful are summarized. Implementing the proposed IBE approach successfully balances the user's right to privacy and the government's right to monitor unlawful messages. The security model's definition describes all entities considered in the model. We demonstrate that the proposed model is secure against IND-CCA of Type I and Type II adversary attacks with their proofs. The defined model served as a framework for stating cryptographic design goals, which were achieved by relying on earlier specified cryptographic building blocks.

The proposed scheme is slightly slower than the previous solution, requiring extra pairing computation and exponential operation during the key issuing algorithm. Nevertheless, since the key issuing process is an algorithm executed once on the server side while the user joins the network, this limitation is not a significant issue. Another limitation is that the scheme assumes that PKG and PKPO never collude with each other. However, the user's Private Key may be compromised if such collusion occurs. Nonetheless, if there is doubt about an unlawful message, the genuine entity from the three entities (PKG, PKPO, and User) may file a case against the malicious entity. Ultimately, the judge may catch a suspected entity out of the three.

In conclusion, the proposed scheme balances privacy and surveillance while ensuring security against adversary attacks. Future research may explore optimizing the key issuing algorithm and addressing the collusion limitation. Our scheme's potential applications in various domains, such as healthcare, finance, and education, warrant further investigation.

## 6.1    Applications and Future Scope

The resulting infrastructure, designed for low computation on the client side and overload shifted to the server side (PKG) on the cloud, is environment-friendly, practical, and instantly ready to use.

- The infrastructure is instantly ready to use, eliminating the need for additional steps once the user joins the network and issues their private key. The encryption and decryption process requires minimal computation cost and the lowest number of public information sharing on the network, providing a user-friendly and efficient solution.

- Moreover, the proposed scheme is environment-friendly, as compared to previous approaches. It balances the user's right to privacy with the government's right to monitor unlawful messages, utilizing a server-side approach to reduce computation on the client side.

- The infrastructure is also practical, as wireless devices increasingly influence daily life, with applications expected to increase daily. The proposed solution shifts the maximum computation operation to the server side, reducing the computations required on the client side. Thus, the proposed solution is applicable for encryption and decryption in low-computation devices like mobiles, laptops, and tablets on the customer side.